\documentclass[10pt, twocolumn]{aastex63}
\usepackage[english]{babel}

\usepackage{bm}
\usepackage{mathtools}
\usepackage{cprotect}
\usepackage{amsmath}

\usepackage{graphicx}

\begin{document}

\title{WONDER: Workload Optimized Network Defined Edge Routing}

\author{Oleg Berzin, Ph.D.}

\affiliation{Fellow, Technology and Architecture, Equinix, Redwood City, CA 94065, USA}
\email{oberzin@equinix.com}

\begin{abstract}
The 5G standards enable cellular network capabilities that significantly improve key network characteristics such as latency, capacity, throughput and reliability, compared to the previous generations of wireless networks. It is, however, clear that in order to achieve these improvements in real network implementations, the supporting physical and logical infrastructure needs to be designed appropriately. The key components of this infrastructure are Radio Access Network, Edge Data Centers, Packet/Optical Interconnection Fabric and Edge Computing. This paper concentrates on the Edge Data Centers, Interconnection and Edge Computing capabilities that target ability to deliver high-performing services on the 5G network by means of intelligent network slicing, traffic routing and coordinated compute workload distribution. We propose new methods of ensuring optimal traffic routing and edge compute workload placement under mobility conditions, subject to application requirements and constraints within a set of interconnected Edge Data Centers, utilizing Segment Routing/IPv6 and software defined control mechanisms. \\
%\newline
\end{abstract}

%\keywords{miscellaneous}

\section{Introduction}
\label{sec:intro}
\subsection*{\raggedright5G}
\label{subsec:5g}
The 5th Generation of Mobile Networks (a.k.a. 5G) represents a dramatic technological inflection point where the cellular wireless network becomes capable of delivering significant improvements in capacity and performance, compared to the previous generations and specifically the most recent one – 4G LTE. In 5G, two key parts of the wireless/mobile network receive major upgrades.
\begin{itemize}
\setlength{\itemsep}{0pt}
\setlength{\parskip}{0pt}
\setlength{\parsep}{0pt}
\item 5G New Radio (5GNR) defines a new air interface structure, new antennae design (massive Multiple Input Multiple Output arrays – mMIMO) and radio transmission methods (beamforming, beam steering).  This enables significant increases in data rates, decreases in the “air interface” latency, as well as the improvements in the capacity (number of connected devices).
\item 5G Core (5GC) defines core control (signaling) and user plane (data) capabilities that make use of cloud native virtualization and enable features such as Control and User Plane Separation (CUPS) and Network Slicing.   This allows for the development of unique services not previously available or difficult to implement in 4G.
\end{itemize}
5G will provide significantly higher throughput than existing 4G networks. Currently 4G LTE is limited to around 150 Mbps. LTE Advanced increases the data rate to 300 Mbps and LTE Advanced Pro to 600 Mbps - 1 Gbps. The 5G downlink speeds can be up to 20 Gbps. 5G can use multiple spectrum options, including low band (sub 1 GHz), mid-band (1-6 GHz) and mmWave (28, 39 GHz). The mmWave spectrum has the largest available contiguous bandwidth capacity ($\approx$1000 MHz) and promises dramatic increases in user data rates. 5G enables advanced air interface formats and transmission scheduling procedures that decrease access latency in the Radio Access Network by a factor of 10 compared to 4G LTE.
\begin{figure}[htp]
    \centering
    \includegraphics[width=5cm]{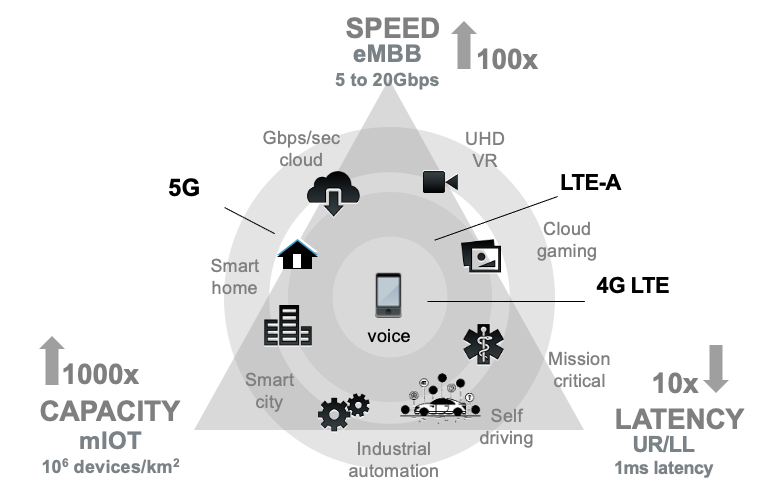}
    \caption{5G Vision}
    \label{fig:5g}
\end{figure}
The full description of 5G Service Based Architecture (SBA) is outside of the scope of this paper. Please refer to \cite{3GPP23501} for the technical specification of SBA. Below we highlight some SBA functions and interfaces that are relevant to the discussion in this document.
\begin{itemize}
\setlength{\itemsep}{0pt}
\setlength{\parskip}{0pt}
\setlength{\parsep}{0pt}
\item UE – User Equipment. These are mobile devices equipped with 5G NR capabilities and used for consumer, business, IoT and vehicular applications.
\item RAN – Radio Access Network. RAN may further be broken down into the Radio Units (RU), Distributed Units (DU) and Centralized Units (CU). The combination of DU and CU is also referred to as gNB. The UE handovers between gNBs are handled using the Xn interface.
\item UPF – User Plane Function. UPF handles Uplink and Downlink data forwarding between the RAN and the Data Networks (DN). The N3 interface is used between gNB and UPF. The N6 interface is used between the UPF and DNs.
\item AMF – Access and Mobility Function. AMF is a Control Plane function responsible for UE registration, session and mobility management.
\item SMF – Session Management Function. SMF is responsible for bearer and session management including the management of UE IPv4 and IPv6 addressing.
\item PCF – Policy Control Function. PCF is responsible for managing and authorizing session parameters for the UE including data rate, QoS and charging.
\item NSSF – Network Slice Selection Function. NSSF is responsible for establishing and managing Network Slicing parameters for the UE sessions including the 5GC and Transport domains. NSSF is described in \cite{3GPP28801}.
\end{itemize}
\begin{figure}[htp]
    \centering
    \includegraphics[width=8.5cm]{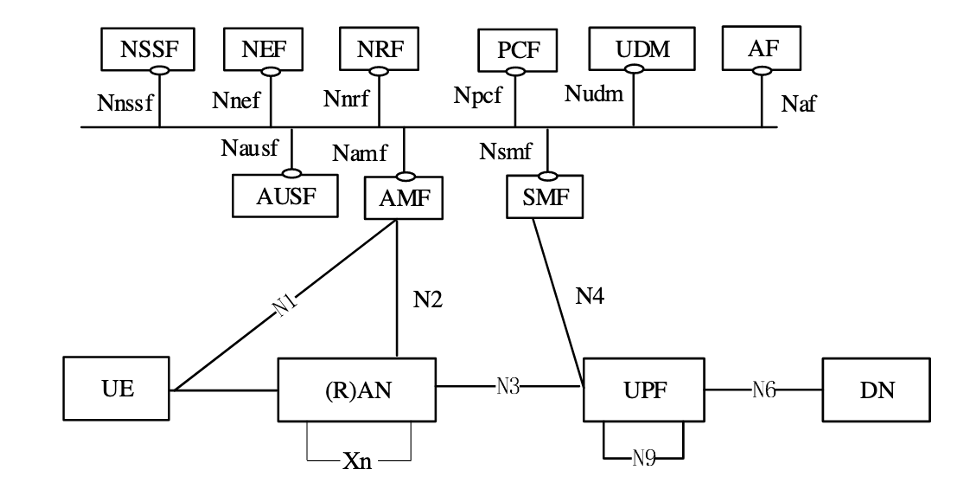}
    \caption{5G Service Based Architecture}
    \label{fig:5g-sba}
\end{figure}

\subsection*{\raggedright {The Edge}}
\label{subsec:edge}
The technological capabilities defined by the standards organizations (e.g., 3GPP, IETF) are the necessary conditions for the development of 5G. However, the standards and protocols are not sufficient on their own. The realization of the promises of 5G depends directly on the availability of the supporting physical infrastructure.

\begin{figure}[htp]
    \centering
    \includegraphics[width=8.5cm]{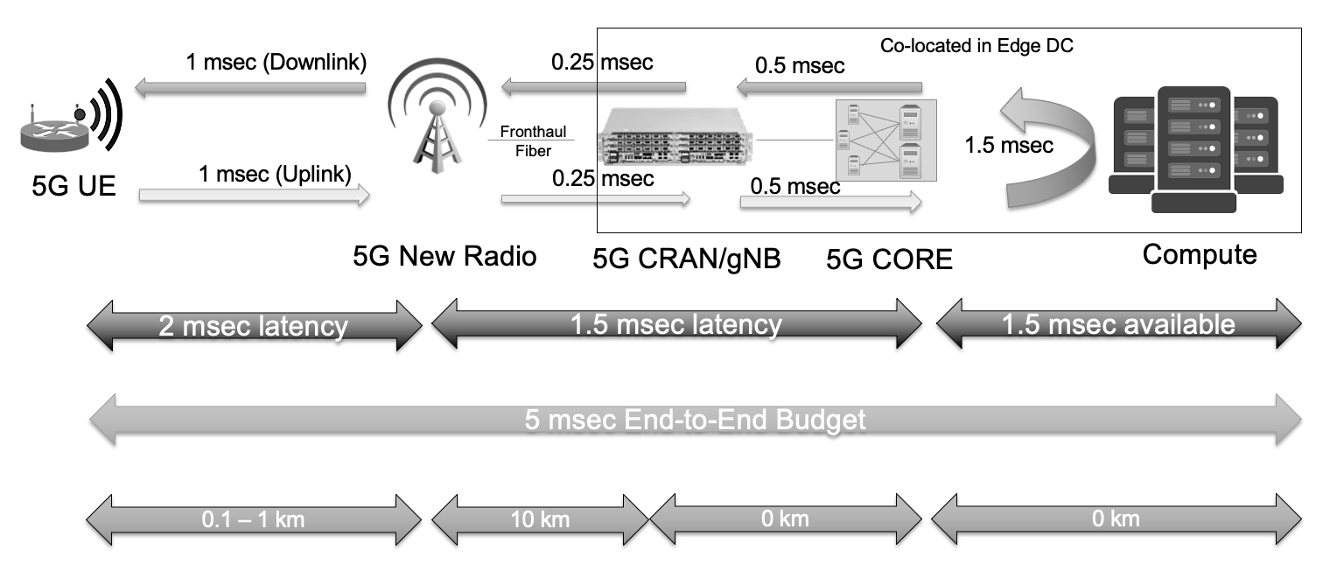}
    \caption{Example of end-to-end latency budget in 5G}
    \label{fig:5g-e2e-l}
\end{figure}

Latency can be used as a very good example to illustrate this point. One of the most intriguing possibilities with 5G is the ability to deliver very low end to end latency. A common example is the 5 milliseconds round-trip device to application latency target. If we look closely at this latency budget, it is not hard to see that to achieve this goal an optimized physical aggregation infrastructure is needed. This is because the 5 millisecond budget includes all radio network, transport and processing delays on the path between the application running on User Equipment (UE) and the application running on the compute/server side. Given that at least 2 milliseconds are expected for the “air interface”, the remaining 3 milliseconds is all that’s left for the transport and the application processing budget. Figure \ref{fig:5g-e2e-l} illustrates the end-to-end latency budget in a 5G network.
As can be seen from Figure \ref{fig:5g-e2e-l}, it is imperative that the 5G Radio Access Network (RAN) equipment, the 5G Core elements and the server computing resources are located in close proximity to each other (i.e. physically co-located) and in a relatively close proximity to the end devices ($\approx$10km). This necessitates the deployment of Edge Data Center (EDC) facilities that enable the infrastructure for the physical aggregation points and for the delivery of one of the fundamental 5G capabilities – the very low end-to-end latency. 

\begin{figure*}[htp]
    \centering
    \includegraphics[width=\textwidth]{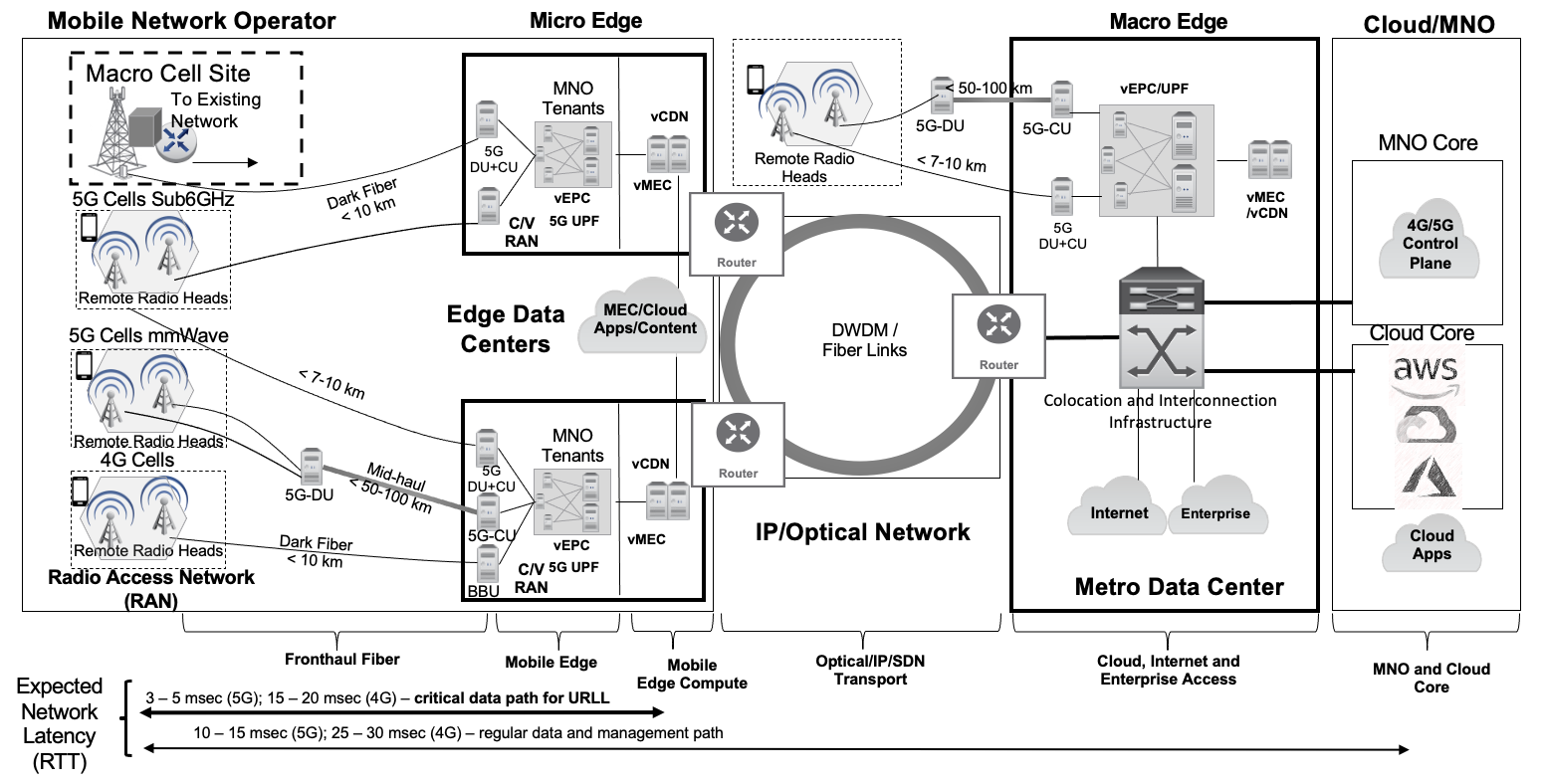}
    \caption{Metro architecture with multiple Micro DCs and Macro DC interconnection}
    \label{fig:ee-arch}
\end{figure*}

Multiple EDCs can be deployed in a given metropolitan area (metro) in order to support multiple functions such as RAN (BBU/DU/CU), 5GC and Multi-access Edge Comuting (MEC) \cite{ETSI-MEC-5G, ETSI-MEC-003}. Due to the need to support higher density of cell sites and the expected area traffic capacity in a given metro, there will be a need for multiple EDCs that are connected to each other as well as to the Macro data center(s) by way of an IP/Optical meshed network. This architecture will allow both the low latency applications that have to make use of MEC resources in EDCs as well as existing applications that rely on the regionally distributed metro infrastructure to connect to the 5G network. An example of this distributed Micro/Macro Edge architecture is shown in Figure \ref{fig:ee-arch}.

\subsection*{\raggedright {The Challenges}}
\label{subsec:thechallenges}
The build-out of a high-performing edge infrastructure to unlock capabilities of the 5G technology is a significant challenge spanning multiple dimensions (e.g. the efficient physical design for EDC, the capital required to deploy an adequate number of EDCs, the real estate procurement to deploy the EDCs, the network deployment to provide the EDCs with connectivity). However, even after the physical and economic issues have been addressed, and the EDC infrastructure has been deployed, there remain additional challenges without solving which, the deployed EDCs may prove less efficient and performing than expected. We classify these challenges into following categories:
\begin{enumerate}
\item Optimized Edge Routing (OER). We expect that the interconnecting network between the EDCs in a metro and between the EDCs and the Macro DC will be an IP/Optical mesh network. The OER expresses the  need of connecting 5G devices to the edge compute and/or macro data center resources in a way that is aware of both the performance requirements (e.g., throughput, latency, survivability) and the network topology (IP and Optical), so that the various traffic flows could be routed according to the application requirements. In addition, there is a need to coordinate the mobile traffic break-out points in the topology with the application workload placement to further optimize the end to end path and performance. The OER challenge is illustrated in Figure \ref{fig:oer}.

\begin{figure}[htp]
    \centering
    \includegraphics[width=8cm]{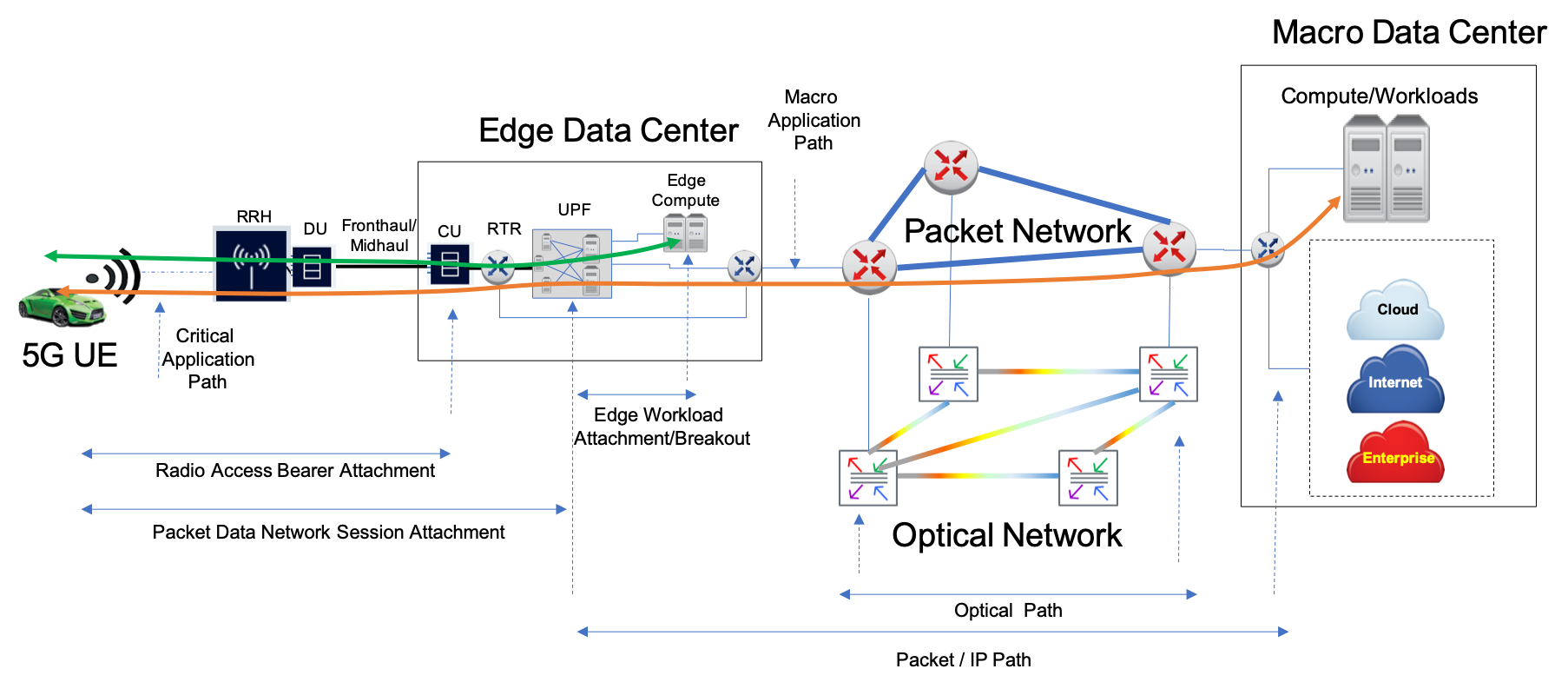}
    \caption{Optimized Edge Routing.}
    \label{fig:oer}
\end{figure}

\item Mobile Edge Routing (MER). Mobility is an inherent property of the 5G network. As devices connect to the 5G network, the mobile data sessions are terminated on the 5G Core elements (e.g. User Plane Function - UPF) located in a given EDC in a metro. As devices (e.g. connected vehicles) move around, they handover from radio cells aggregated at a given EDC to radio cells aggregated at another EDC. While the radio handover takes place, the mobile session path may still be directed to the original 5G Core element (UPF) in the EDC that was used at the start of the session. This may create a suboptimal network path between the device’s new point of attachment in the radio network and the application workload that may still be located in the original EDC. Thus, there is a need to find the best path, subject to end-to-end constraints (e.g. latency), between the EDC where the device is attached at the radio layer and the “remote” EDC where the mobile session is terminated and the application workload for the mobile session is executing. The MER challenge is illustrated in Figure \ref{fig:mer}.

\begin{figure}[htp]
    \centering
    \includegraphics[width=8cm]{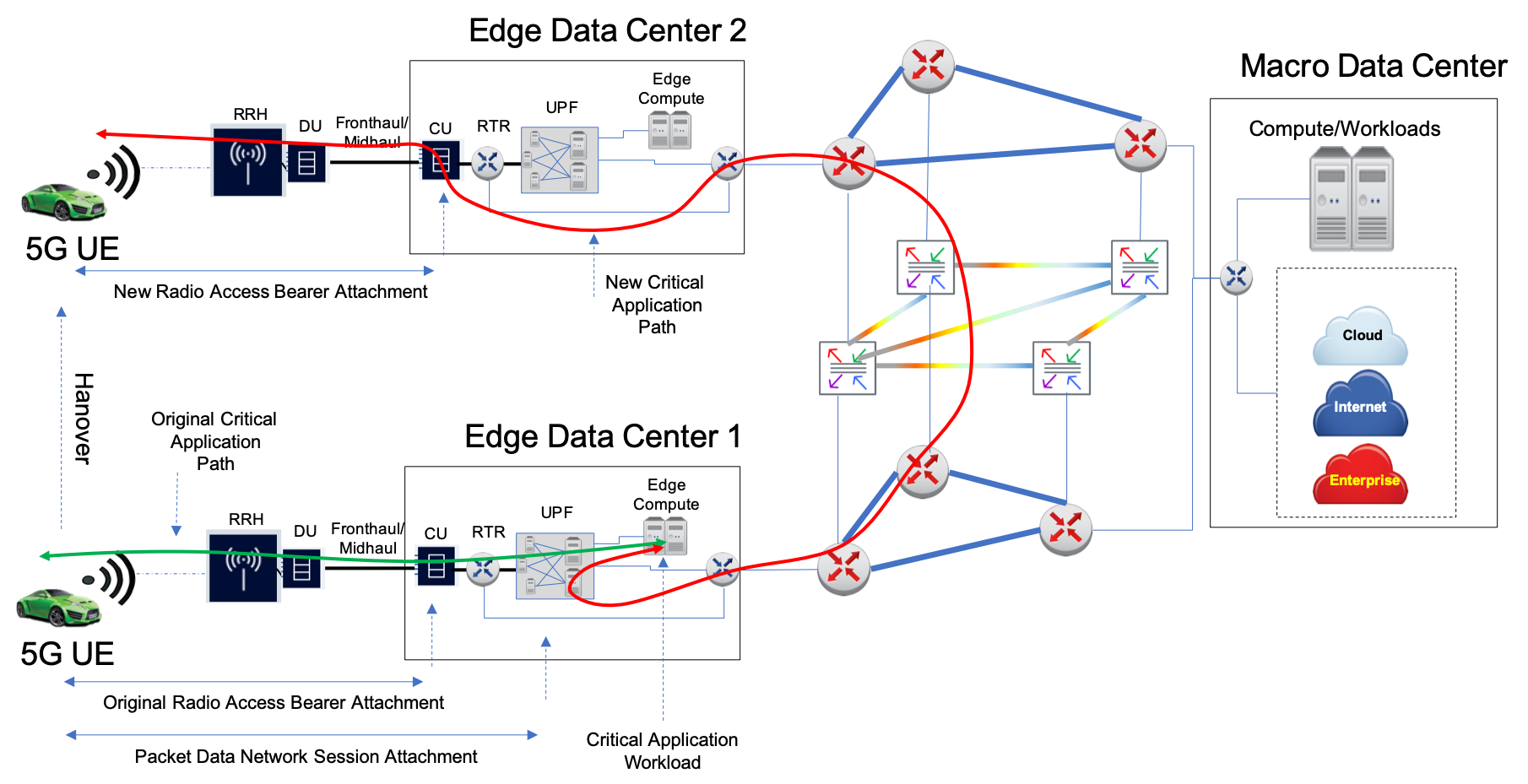}
    \caption{Mobile Edge Routing.}
    \label{fig:mer}
\end{figure}

\item Application Workload Routing (AWR). Building on the Mobile Edge Routing issue, it is conceivable to envision that the critical application traffic may be negatively affected by the additional network distance if both the mobile session and the application workload are still anchored at the original EDC and are effectively lagging behind the moving user. It is therefore desirable to “break-out” the critical application traffic locally at the current EDC. One critical issue with this local break-out is the ability to maintain the data connectivity to the UE while also maintaining the optimal traffic path to the UE. In addition, for critical applications, this local break-out only makes sense if the application workload is also moved or replicated to the current EDC. Thus, there is a need to coordinate the decision for the local break-out of the mobile traffic with the ability to move or replicate the application workload along the most optimal path to the compute resources local to the break-out point. The AWR challenge is shown in Figure \ref{fig:awr}.

\begin{figure}[htp]
    \centering
    \includegraphics[width=8cm]{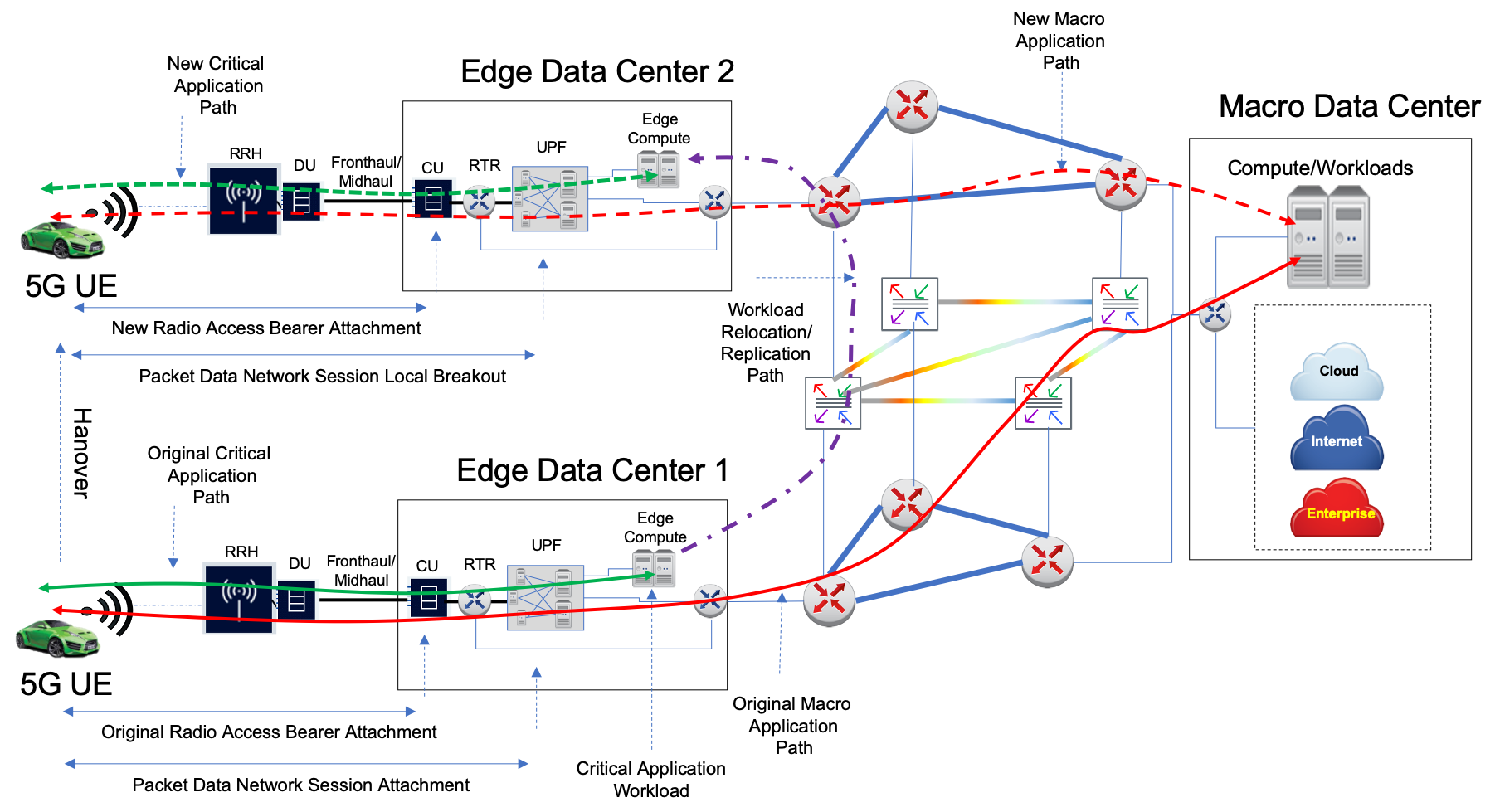}
    \caption{Application Workload Routing.}
    \label{fig:awr}
\end{figure}

\item Application Workload Inter-working (AWI). The 5G network deployment will be provided by multiple carriers (incumbent and new). Therefore, we should expect various 5G UE OEMs make use of different 5G providers and their network deployments. For example, a vehicle produced by Manufacturer A may be using 5G service from Provider 1, while another vehicle by Manufacturer B may prefer to use 5G service from Provider 2. It is natural to expect that both vehicles would need to communicate with a common application and share/consume critical application information regardless of the 5G service/network provider and within the required performance constraints (e.g. latency, throughput, reliability). AWI would ensure that traffic to/from mobile devices served by different networks can be routed in an optimal way to a common application workload while preserving the capabilities enabled by the OER, MER and AWR. In addition, AWI would enable replication of critical application workloads to the appropriate traffic break-out points in the new EDC along the most optimal inter-EDC network path. The AWI challenge illustrated in Figure \ref{fig:awi}.

\begin{figure}[htp]
    \centering
    \includegraphics[width=8cm]{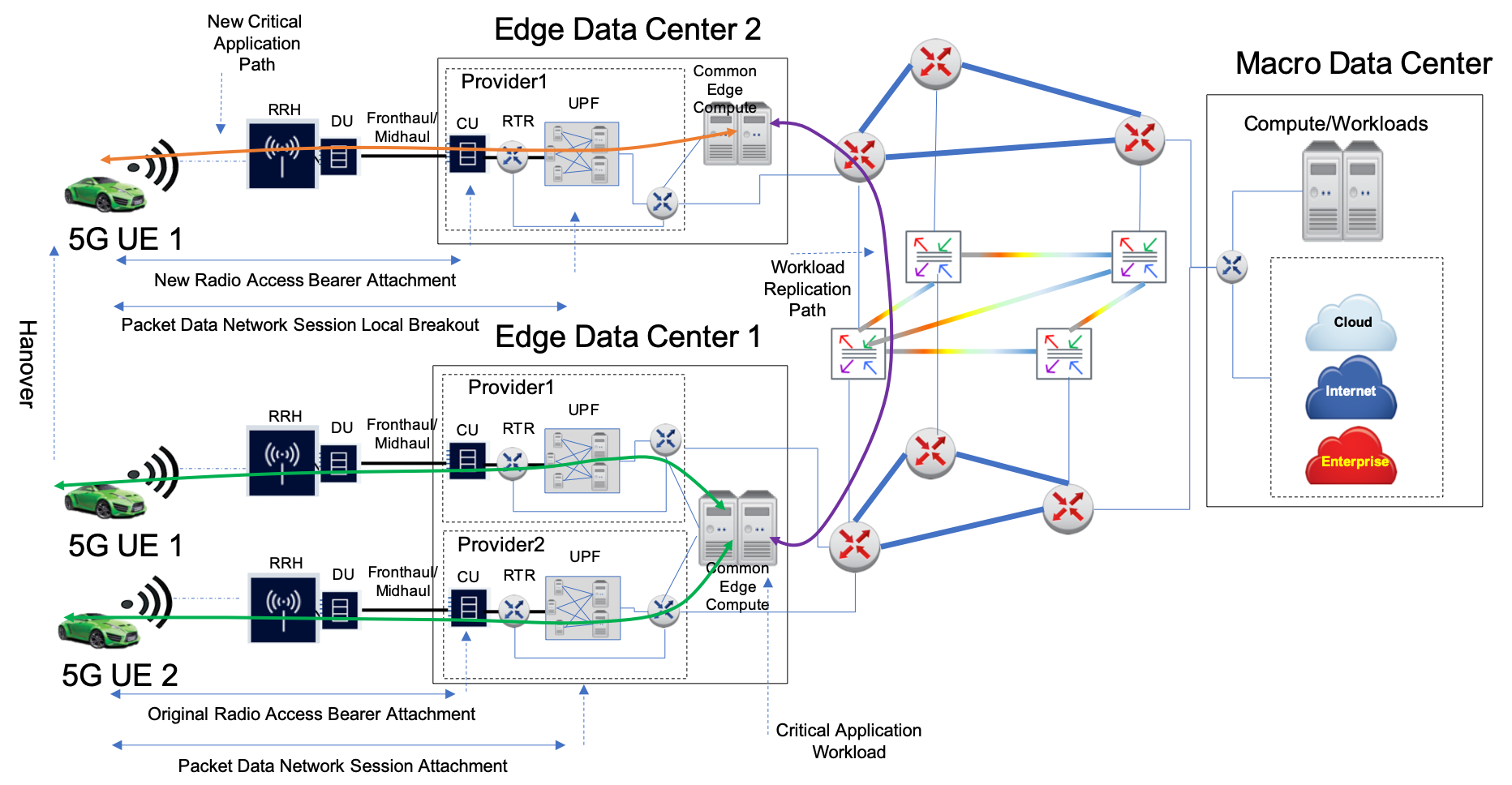}
    \caption{Application Workload Inter-working.}
    \label{fig:awi}
\end{figure}
\end{enumerate}

In addition to the challenges described above, in \cite{AECC-Edge} the authors describe key issues related to the automotive/vehicular use cases that are relevant to the discussion in this paper:
\begin{itemize}
\item Edge Data Offloading. This issue brings up questions on specific methods that can be invoked in the 4G and 5G networks to enable offloading or routing of certain relevant application traffic between the UE (in a vehicle) and the Edge Compute server/resource. This issue is also relevant to routing the UL and DL traffic between the UE and the Edge Compute server while handling the mobile session continuity and IP addressing/reachability.
\item Edge Compute Server Selection. This issue poses a question of what methods can be used to appropriately select Edge Compute resources for a specific application required by the UE while satisfying performance requirements.
\item UE or Vehicle System Reachability. This issue brings up questions of how to reach the UE in an unsolicited manner (i.e. without the UE explicitly sending UL traffic). This issue is not directly related to the discussion in this document.
\end{itemize}

The purpose of this paper is to propose solutions to the challenges described above in the context of the 5G network deployment utilizing a set of interconnected Edge Data Centers. 

\section{Proposed Approaches}
\label{sec:approaches}

\subsection*{\raggedright{Optimized Edge Routing}}
\label{subsec:oer}
We first address the issue of being able to route mobile data sessions (PDN sessions and IP sessions) in accordance with the requirements of the applications carried as traffic along the path that the data sessions provide. First, consider a requirement for two example application classes that need to be supported to/from the 5G enabled vehicle (note that the vehicular scenario and the two classes are used as an example only and that the proposed approach is not restricted to two classes or the vehicular use case):
\begin{itemize}
\item The low latency class, where the application providing critical services to the vehicle must be located as close as possible to the vehicle’s position. An example of such application is a Cellular Vehicle to Infrastructure (C-V2I) congestion avoidance or traffic look-ahead capability, or a Vulnerable Road User (VRU) application that lets the vehicle know of an imminent collision with a pedestrian or a bicyclist or a motorbike user. For the purposes of this paper we refer to this class as "Class 0".
\item The high data rate class, where the applications roughly fall into the infotainment category and range from video content delivery to the passengers, to general internet access, to other services such as intelligent parking and high-definition maps. For the purposes of this paper we refer to this class as "Class 1".
\end{itemize}
We assume that Class 0 traffic must be routed from the RAN and terminated on an application workload (e.g. a Kubernetes \cite{K8S} container) that runs the corresponding service instance(s). The ideal physical location for such a container is locally within the Mobile\footnote{In the context of this paper the terms Multi-access and Mobile are used interchangeably and with equal meaning to describe MEC resources} Edge Compute (MEC) infrastructure in the same Edge Data Center (EDC) which hosts the RAN/5GC elements that services the mobile session providing connectivity for the Class 0 traffic.

Each Traffic Class may have a set of parameters associated with its definition, such as (not limited to) latency bound, peak data rates, maximum error rate, reliability and restoration requirements, packet layer QoS, 5G QFI, etc. An example Traffic Class definition is shown in Table \ref{table:1}.
\begin{table}[htp]
\begin{center}
\footnotesize
\begin{tabular}{||p{0.7cm}|p{0.9cm}|p{0.7cm}|p{1.7cm}|p{1cm}| p{1.7cm}||}
 \hline\hline
 Class ID & Latency Bound (RTT) & Peak Data Rate & Resiliency & Max Error Rate & QoS \\ [0.5ex]
 \hline \hline
 0 & 7 ms & 1 Gbps & Protected 50 ms restoration & 0.01\% & 5G QFI, MPLS EXP, IPv6 Flow Label
\\ 
 \hline
 1 & 20 ms & 5 Gbps & Unprotected, & 0.1\% & 5G QFI, MPLS EXP, IPv6 Flow Label \\ [0.5ex]
 \hline\hline
\end{tabular}
\caption{Example Traffic Class Definitions.}
\label{table:1}
\end{center}
\end{table}

The standardized Traffic Class definitions may be agreed to between the Mobile Network Operators and the providers of the inter-EDC interconnection services.

Recognizing the possibility that not all EDCs that are equipped with the RAN functions and are capable of accepting wireless connections from devices, are also equipped with the Edge Compute functions (for example, some EDCs could be smaller Centralized RAN (CRAN) hubs without the Edge Compute servers), we introduce a concept of a MEC Domain. 

\begin{figure}[htp]
    \centering
    \includegraphics[width=8cm]{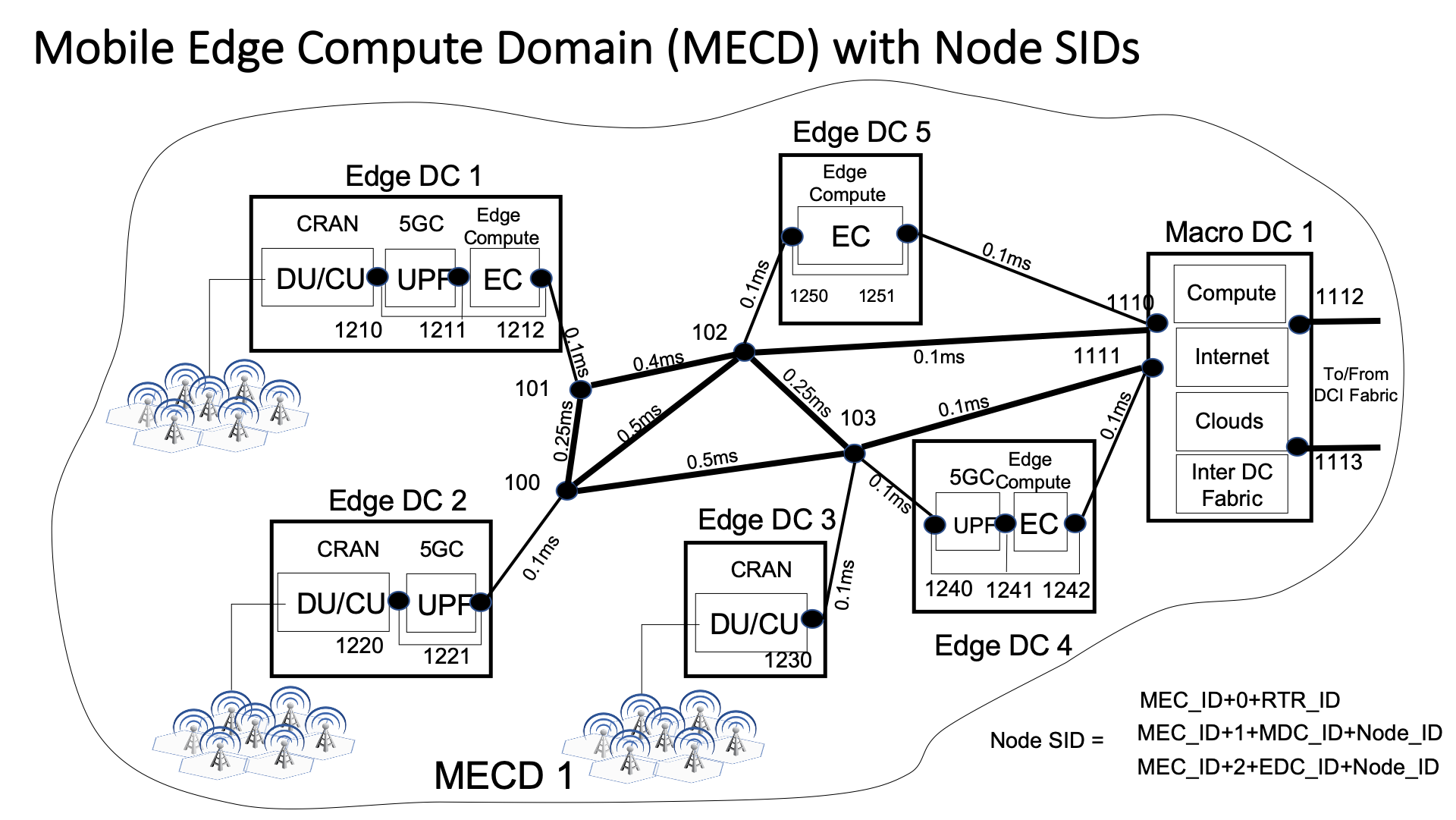}
    \caption{Mobile Edge Computing Domain.}
    \label{fig:mecd1}
\end{figure}

\begin{figure}[htp]
    \centering
    \includegraphics[width=8cm]{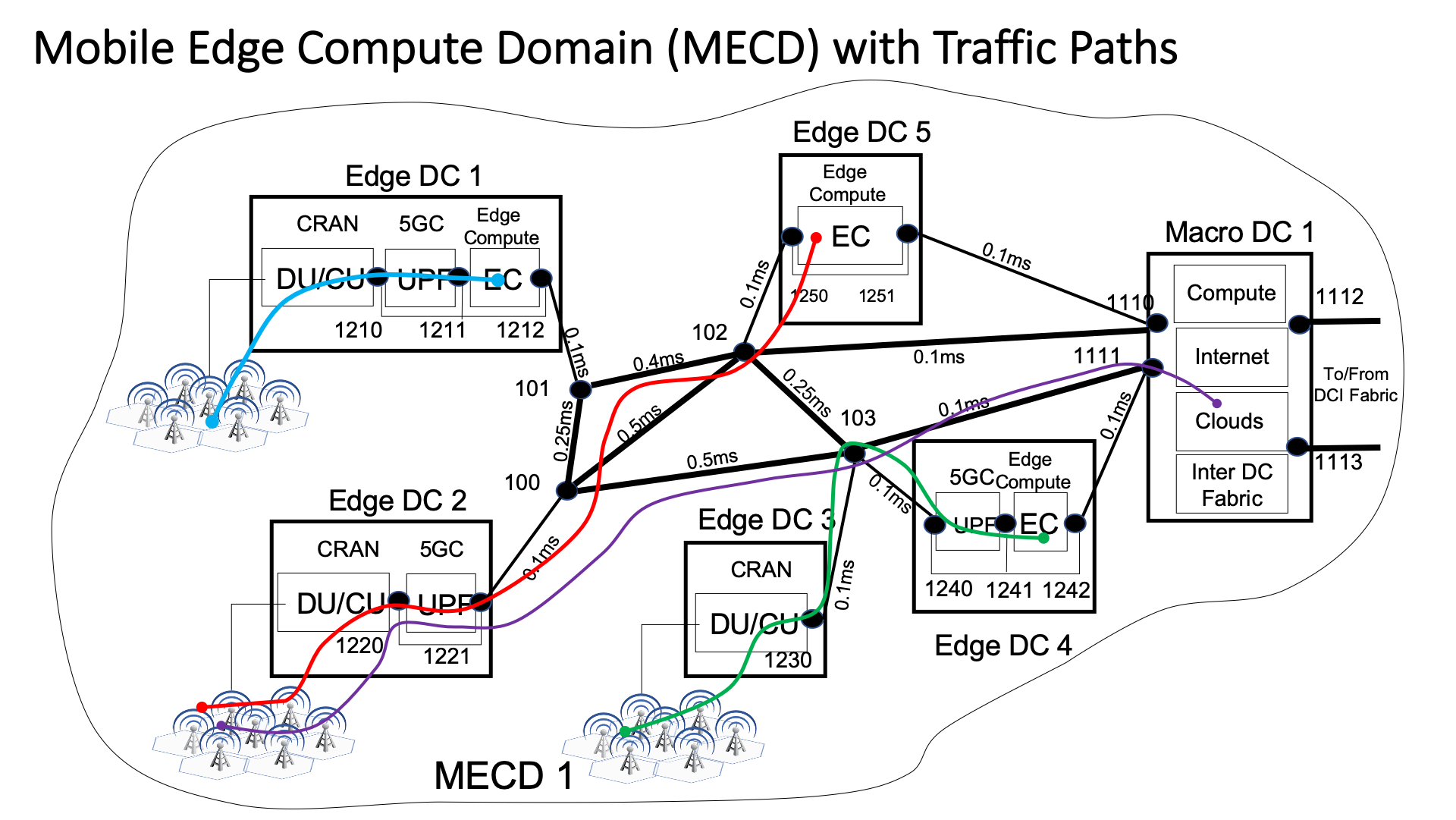}
    \caption{Traffic Paths in a Mobile Edge Computing Domain.}
    \label{fig:mecd2}
\end{figure}

A MEC Domain (MECD) is a set of EDCs that define a certain geographic zone which is expected to provide a specific upper bound for a latency budget (e.g. 5-7 ms of round-trip time from a device attached to the 5G RAN to an application workload running on any server within the corresponding MECD). An example of MECD is shown in Figure \ref{fig:mecd1}. At least one EDC in a MECD must be equipped with Edge Compute (EC) resources.
Further, we specifically associate a set of Segment Routing (SR) \cite{SR} labels with all possible traffic paths within the MECD. Some of the possibilities include:
\begin{itemize}
\item A path from UE to the combination of CU, UPF and EC located in the same physical EDC. For example, Label Stack from Figure \ref{fig:mecd1} (1210, 1211, 1212) with the path shown in blue in Figure \ref{fig:mecd2}.
\item A path from UE to CU in one EDC to UPF and EC that are both located in a different physical EDC. For example, Label Stack from Figure \ref{fig:mecd1} (1230, 103, 1240, 1241) with the path shown in green in Figure \ref{fig:mecd1}.
\item A path from UE to CU and UPF in one EDC and to the EC in another EDC. For example, Label Stack from Figure \ref{fig:mecd1} (1220, 1221, 100, 102, 1250) with the path shown in red in Figure \ref{fig:mecd2}.
\end{itemize}
Note that the SR labels may be implemented using SRv6 \cite{SRv6}, \cite{SRv6-RFC8754}, in which case the actual data plane for the transport of traffic is based on IPv6 rather than MPLS, the SR labels themselves may be formatted as 128-bit IPv6 addresses, but the Segment Routing logic is still driven by the SR label stack expressed as the SR Header inside the IPv6 packet.

The use of MPLS based approaches to enhance traffic routing in the mobile network (without the use of GTP-U) has been proposed in multiple publications, including \cite{MLBN, HMLBN-ARCH, HMLBN-MOD, HMLBN-IEEE, MLBN-DRAFT, HMLBN-Drexel}. Segment Routing furthers these methods by enabling the architecture where the distribution of the label information can be handled by the Control Plane capable of computing necessary paths to meet the traffic class requirements as well as performing traffic engineering using source routing based on SR labels. SRv6 takes this to the next level, where the requirement for the MPLS forwarding plane is lifted and is replaced by IPv6 but the Control Plane and traffic engineering aspects are retained by using Segment Routing processing.

In general, the path matrix may be constructed with an appropriate degree of granularity, listing all possible paths linking all possible source-destination pairs within the MECD. Each pair and the associated paths may then be further classified in accordance with their definitions as shown in Table \ref{table:1}. One possible coarse degree of granularity is EDC to EDC within the MECD. Another possibility is any CU to any UPF and/or any UPF to any EC.

We propose that the MECD SR Label Set that describes these paths is computed and stored in the software defined entity referred to as the OER Controller.

We further propose that the OER Controller (OERC) is aware and capable of correlating the network fabric path information based on the link and routing metrics between the packet and optical domains. In other words, the OER Controller should be capable of identifying a packet path based on the optical path parameters (such as span latencies and capacities as well as the protection state). Note that the link delays in Figure \ref{fig:mecd2} are shown as optical network delays. The requirement to coordinate the optical path parameters with the packet routing paths means that inter-working between the optical and packet controllers should exist under the OERC framework. 
 
\begin{figure}[htp]
    \centering
    \includegraphics[width=8.5cm]{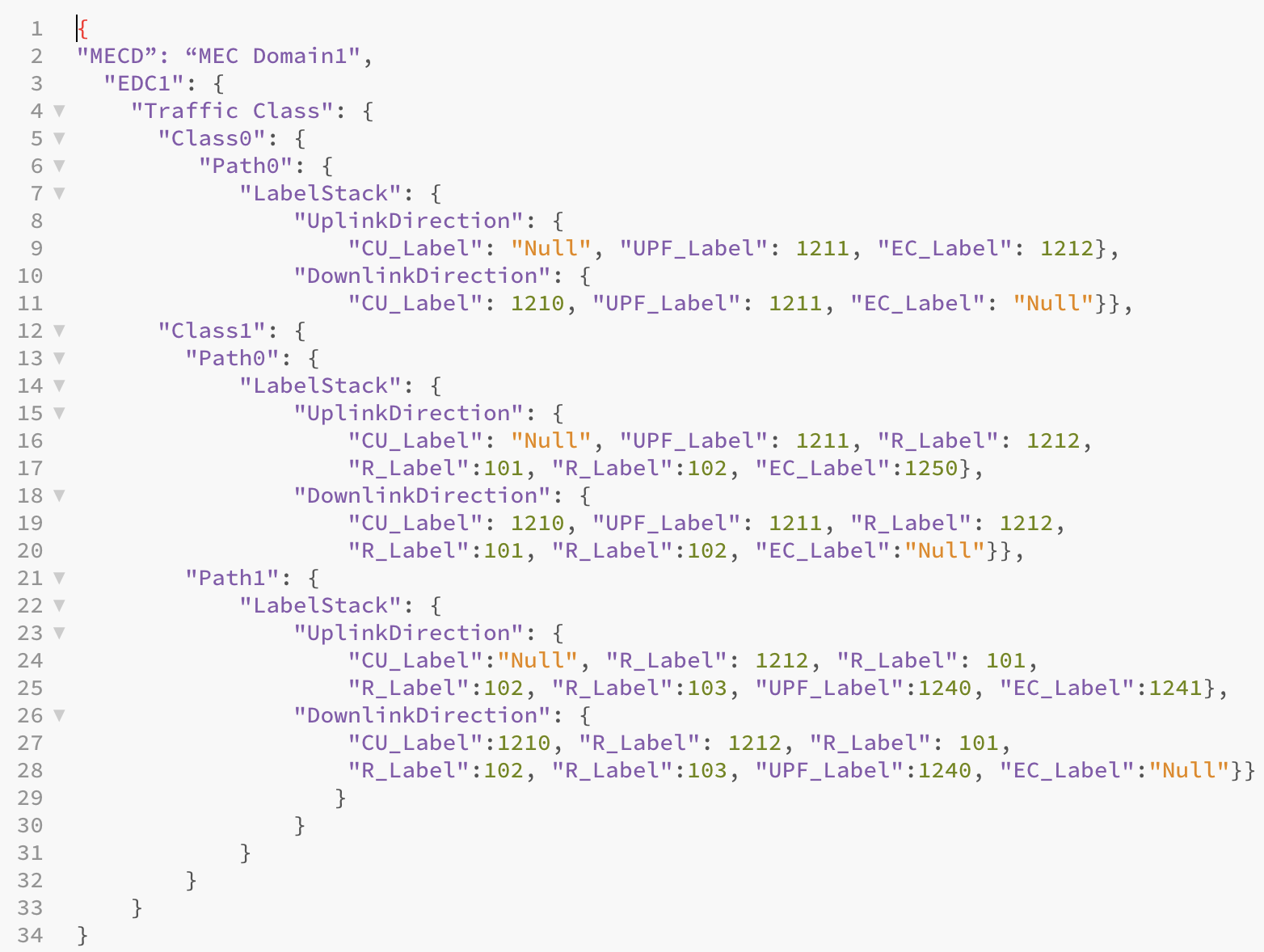}
    \caption{Traffic Class definition example for OER Controller.}
    \label{fig:mecd-path-json}
\end{figure}

In addition, we propose that the traffic classes are defined in the OERC in such a way as to provide a means of identifying requests for routing the traffic of a certain class so that the path can be contained within the expected performance bounds and within the MECD. For example, a mapping structure could be implemented in the OERC as shown in Figure \ref{fig:mecd-path-json}.

\begin{figure}[htp]
    \centering
    \includegraphics[width=8cm]{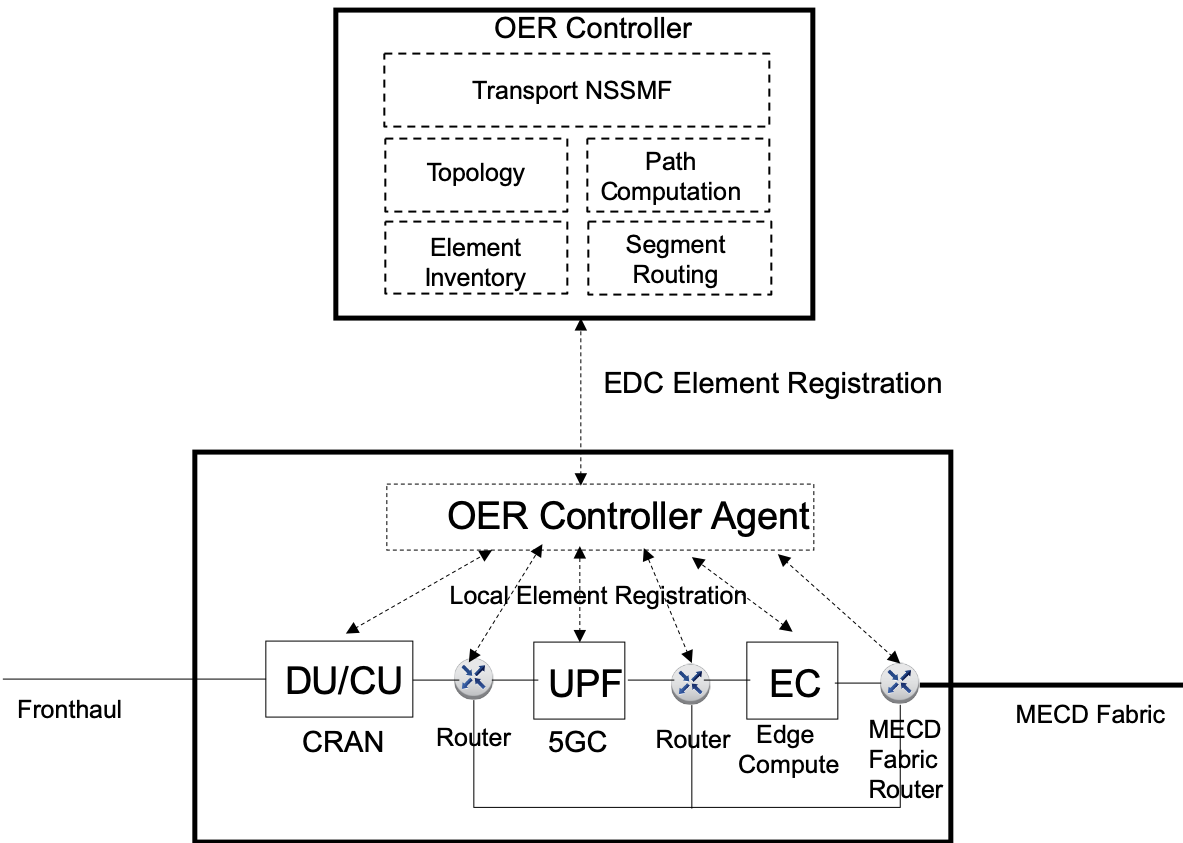}
    \caption{Edge DC Element Registration Architecture.}
    \label{fig:edc-reg}
\end{figure}

\begin{figure}[htp]
    \centering
    \includegraphics[width=6cm]{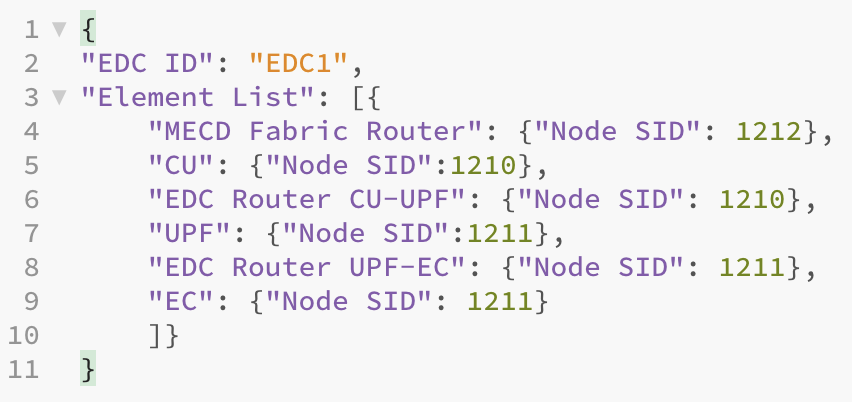}
    \caption{EDC Element Registration Record.}
    \label{fig:mecd-reg-json}
\end{figure}

The MECD paths “Path j” are calculated by the OERC based on the element registration information provisioned during the network design/implementation. For example, all CUs, Routers, UPFs and EC resources within a given EDC across all EDCs in the MECD may be registered with the OERC for the domain. This could be achieved in a hierarchical manner by enabling the EDC-level OERC Agents. Similarly, all transport router and optical nodes may register their Node SIDs with the OERC. The example EDC element registration architecture is shown in Figure \ref{fig:edc-reg}.
The structure of the registration record may be based on the EDC identification, the identification of functional elements with the EDC and the SR Segment IDs for the elements that are responsible for traffic routing and forwarding. We propose that the list of such elements includes CU, EDC Routers, UPFs. Note that the SR labels may be implemented as SRv6 128-bit labels if SRv6 is used for user plane traffic forwarding. The EDC OER Registration record may look as shown in Figure \ref{fig:mecd-reg-json}.

\begin{figure}[htp]
    \centering
    \includegraphics[width=8cm]{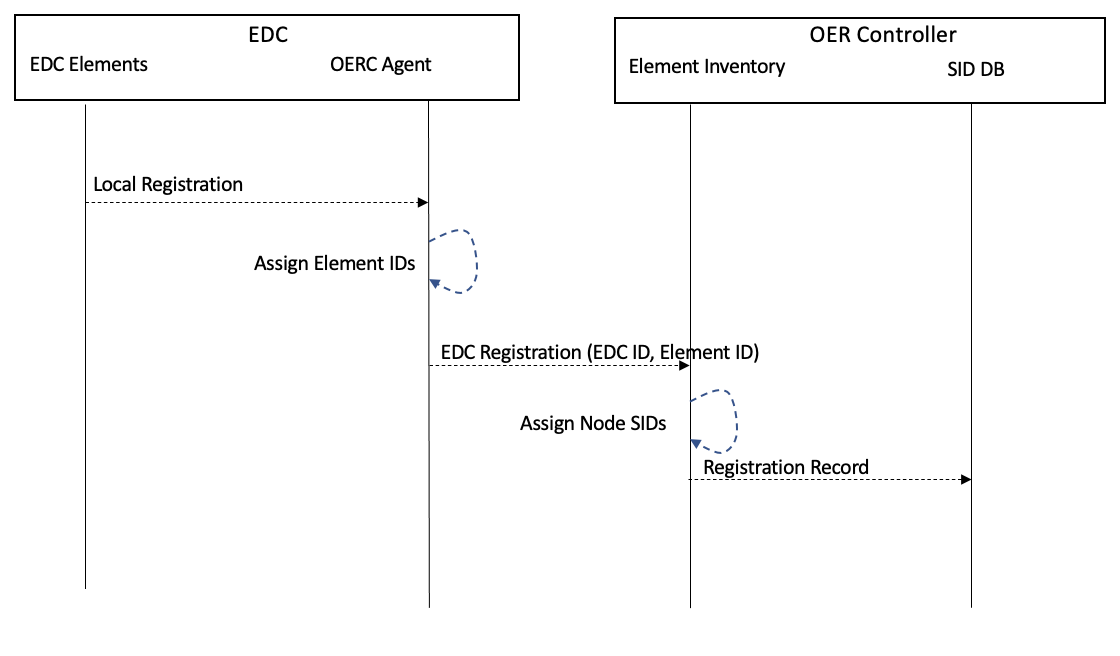}
    \caption{Edge DC Element registration procedure.}
    \label{fig:edc-reg-proc}
\end{figure}
The effect of this structure is that by the time the network of EDCs is deployed in a given area and the EDCs are equipped with particular network elements, the OERC for this area will have had a full inventory of the functions installed in all the EDCs under its control along with the definitions of traffic classes mapped to the associated possible network paths that satisfy the traffic class requirements, with the network paths expressed as sets of SR/SRv6 labels. An example EDC element registration procedure is shown in Figure \ref{fig:edc-reg-proc}.

Note that specific algorithms for computing the paths for the traffic classes may be based on a variety of existing well-known techniques or any other applicable methods.
\begin{figure}[htp]
    \centering
    \includegraphics[width=8cm]{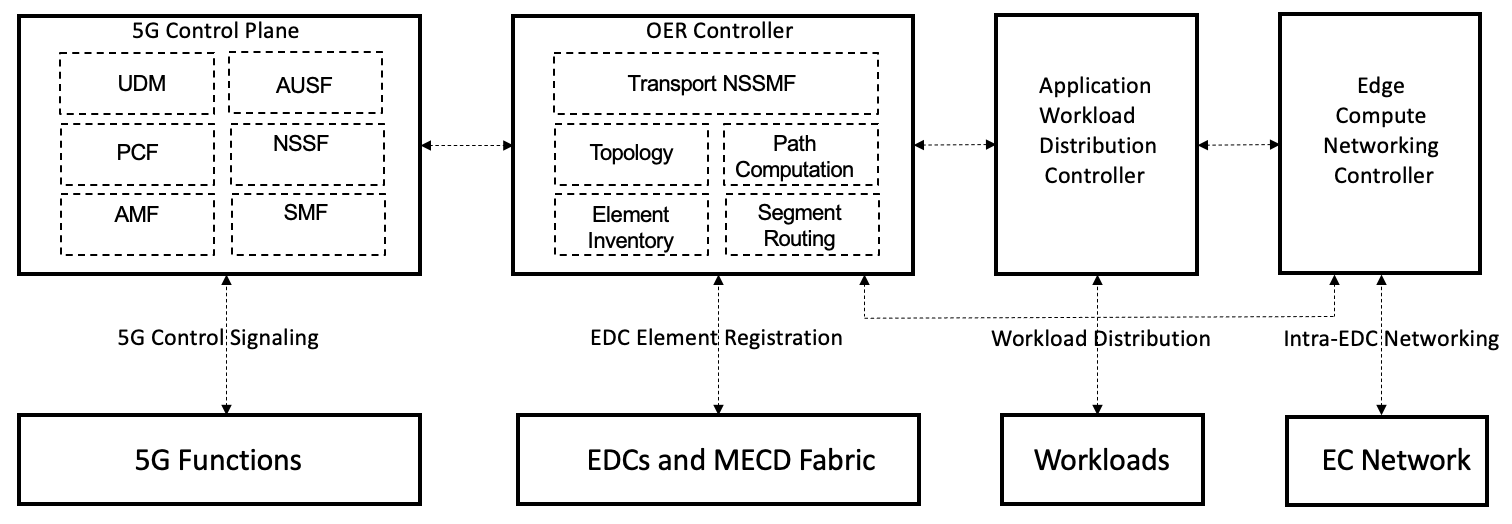}
    \caption{Controller Interactions.}
    \label{fig:controllers}
\end{figure}
The high-level controller interactions are shown in Figure \ref{fig:controllers}. The 5G Control Plane (5GCP) handles all signaling related to providing connectivity and services to 5G devices or User Equipment (UE). One of the 5GCP functions is NSSF - Network Slice Selection Function. We propose that NSSF be the entity that interacts with the OER Controller. 

The OERC is responsible for the EDC element registration (as discussed above), MECD topology discovery, path computations and SR/SRv6 label assignments. The OERC interacts with 5GCP, MECD Fabric nodes, Application Workload Distribution Controller (AWDC) and the Edge Compute Networking Controllers within the EDCs.

The AWDC \footnote{The implementation od AWDC may be based on Kubernetes.} is responsible for workload scheduling and distribution within the Edge Computing environments in the EDCs.

The Edge Compute Networking Controllers (ECNC) are responsible for all aspects of networking within the Edge Computing environments, including the support for SR/SRv6. Note that SR/SRv6 may be supported directly the EC servers (for example, on SmartNICs). This allows for end-to-end SR/SRv6 traffic processing regardless of the UE and EC workload IP addressing.

\begin{figure}
    \centering
    \includegraphics[width=8.5cm]{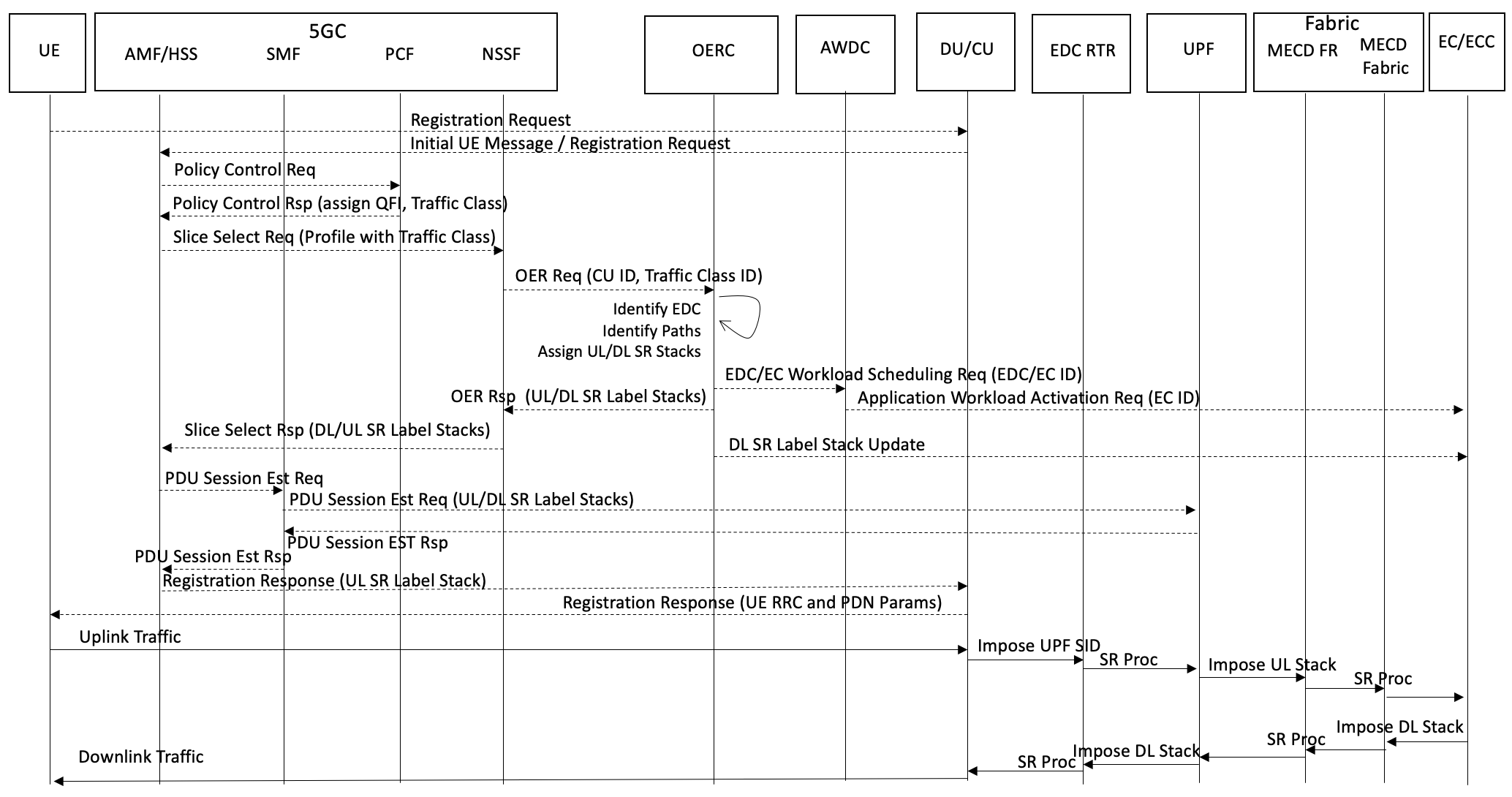}
    \caption{OER End to End Signaling Procedure.}
    \label{fig:oer-sig}
\end{figure}

\begin{figure}
    \centering
    \includegraphics[width=8.5cm]{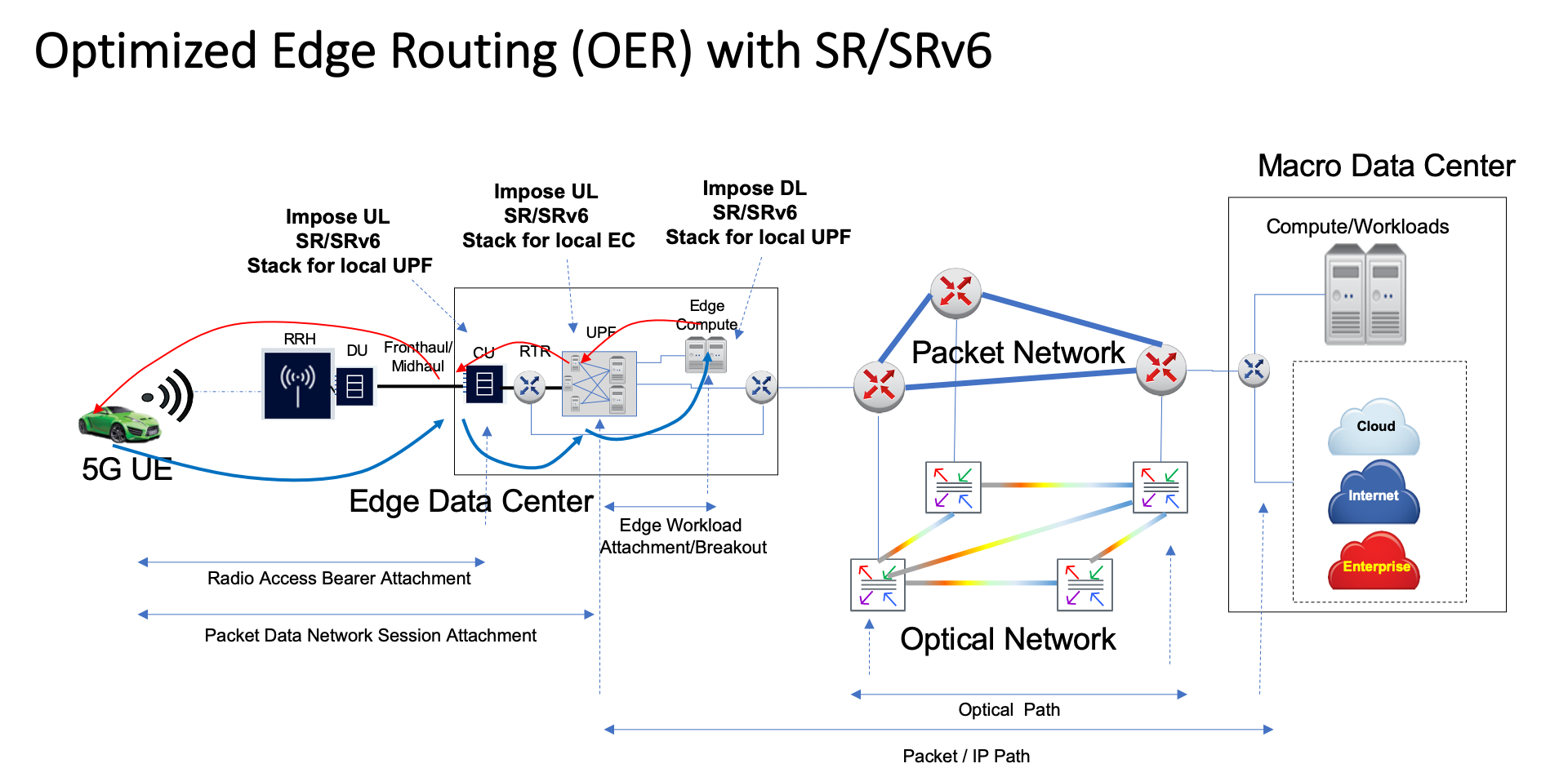}
    \caption{OER traffic forwarding using SR/SRv6.}
    \label{fig:oer-srv6}
\end{figure}

Given this setup, we propose the following procedure \footnote{The proposed signaling procedures and call flows are intended to illustrate a general approach to enabling the OER capability} for enabling Optimized Edge Routing (OER) within a given MEC Domain:

\begin{enumerate}
\item The UE performs registration, authentication and attachment procedures to connect to the 5G \footnote{This paper does not intend to replicate precise 5G signaling procedures and shows generalized call flows roughly following 5G specifications} network.
\item The 5G Control Plane (5GCP) performs all appropriate functions for UE attachment (the below is not the exhaustive list):
\begin{itemize}
    \item AMF registration procedures
    \item UDM Subscriber Identification/Authentication
    \item PCF policy control, 5G QoS Flow identification (QFI) and parameter selection, including Traffic Class ID \footnote{The Traffic Class ID in PCF parameters is not currently specified in 5G procedures}
    \item AMF/SMF interactions and selection of DU, CU and initial UPF for the UE, 5G QoS Flow ID (QFI) assignment for the Traffic Class required for the UE
\end{itemize}
    \item NSSF (Network Slice Selection Function) sends an OER Request \footnote{The OER Request / Response signaling is proposed as new functionality} to the OERC with:
\begin{itemize}
    \item Traffic Class ID, CU ID, UPF ID, QFI. Notes:
a)	The element IDs for CU and UPF may be correlated to the MECD element IDs registered by the EDC OERC Agent to the MECD OERC. b) The CU ID can be used to identify the source EDC for the path selection as well as for the identification of where to schedule the EC workload.
\end{itemize}
\item The OERC consults the OER Path Selection Database for the MECD and identifies a valid path record within the best set of constraints to satisfy the traffic class requirements received from the NSSF.
\item The OERC performs the following:
\begin{itemize}
    \item Responds to NSSF with the OER Response containing the SR label stacks for the mobile elements, CU and UPF, computed for the UE session. Notes: 
    a) The CU and UPF are required to support SR and/or SRv6. b) The ability of CU to use SR/SRv6 is intended to deliver the best network path starting as deep in the mobile network as the CU also known as a gNB. c) The NSSF is expected to communicate the information received from OERC, to the AMF. The AMF may replace the original UPF selected for the UE during the initial node selection procedure with the UPF corresponding to the UPF Node SID received from the OERC based on the path computation results. This is a way to influence the UPF node selection based on the path routing within the MECD. In other words, this is how the workload routing/scheduling can influence where the mobile traffic can terminate optimally relative to the workload location. d) The AMF signals the OER SR label stack to SMF and CU. The SMF signals the label stack information along with the mobile session parameters for the UE to the UPF. The SR or SRv6 labels may be used by the CU and UPF in the same way as GTP-U TEIDs are used to manage the packet switching in GTP-U.
    \item The SR label stack communicated from the OERC to NSSF/AMF/SMF/CU/UPF is used to handle the traffic in the Uplink direction. Notes: a)The CU is expected to use the UPF Node SID label for packets on the N3 interface toward the UPF. The CU may use SRv6 for the same function. b) The UPF is expected to use the remaining SR/SRv6 label stack in the packets on the N6 interface toward the MECD Fabric Router in the EDC. c) The MECD Fabric router is expected to process the SR/SRv6 label stack and send the packets upstream toward the MECD Fabric.
    \item Communicates the EC node identity to the Application Workload Distribution Controller (AWDC). This is expected to allow the AWDC to activate application specific workload worker(s) on the compute infrastructure of the selected EC node. This capability is used to for the Application Workload Routing (see section \ref{subsec:awr}.
    \item Communicates the Downlink SR/SRv6 label stack to the selected EC node or to the EC Network Controller (this is the SDN controller responsible for the networking in the EC compute pod). Notes: a) The EC node networking stack (running on the compute OS itself or on the SmartNIC, or on the EC networking fabric) is expected to support Segment Routing or Segment Routing IPv6. b) The EC networking stack is expected to insert the received SR/SRv6 label stack into packets going in the Downlink direct toward the MECD Fabric. c) The MECD Fabric nodes, including the MECD Fabric Router in the EDC, are expected to process the packets in accordance with the SR/SRv6 label stack carried in the packets and deliver packets to the corresponding UPF in the associated EDC. d) The UPF is expected to receive SR/SRv6 packets on the N6 interface, process the label stack, insert the CU Node SID label and send the packets on the N3 interface toward the CU. e) The CU is expected to receive the SR/SRv6 packets, discard the label stack and process the packets in accordance with the downlink procedures toward the DU/RRH.
\end{itemize}
\end{enumerate}
The effect of the above procedure is the programming of the best (a.k.a OER) path from the UE to the Edge Compute workload based on the traffic class requirements of the corresponding application. The example signaling sequence and the high-level traffic flow are shown in Figures \ref{fig:oer-sig} and \ref{fig:oer-srv6} respectively.

\begin{figure}
    \centering
    \includegraphics[width=8.5cm]{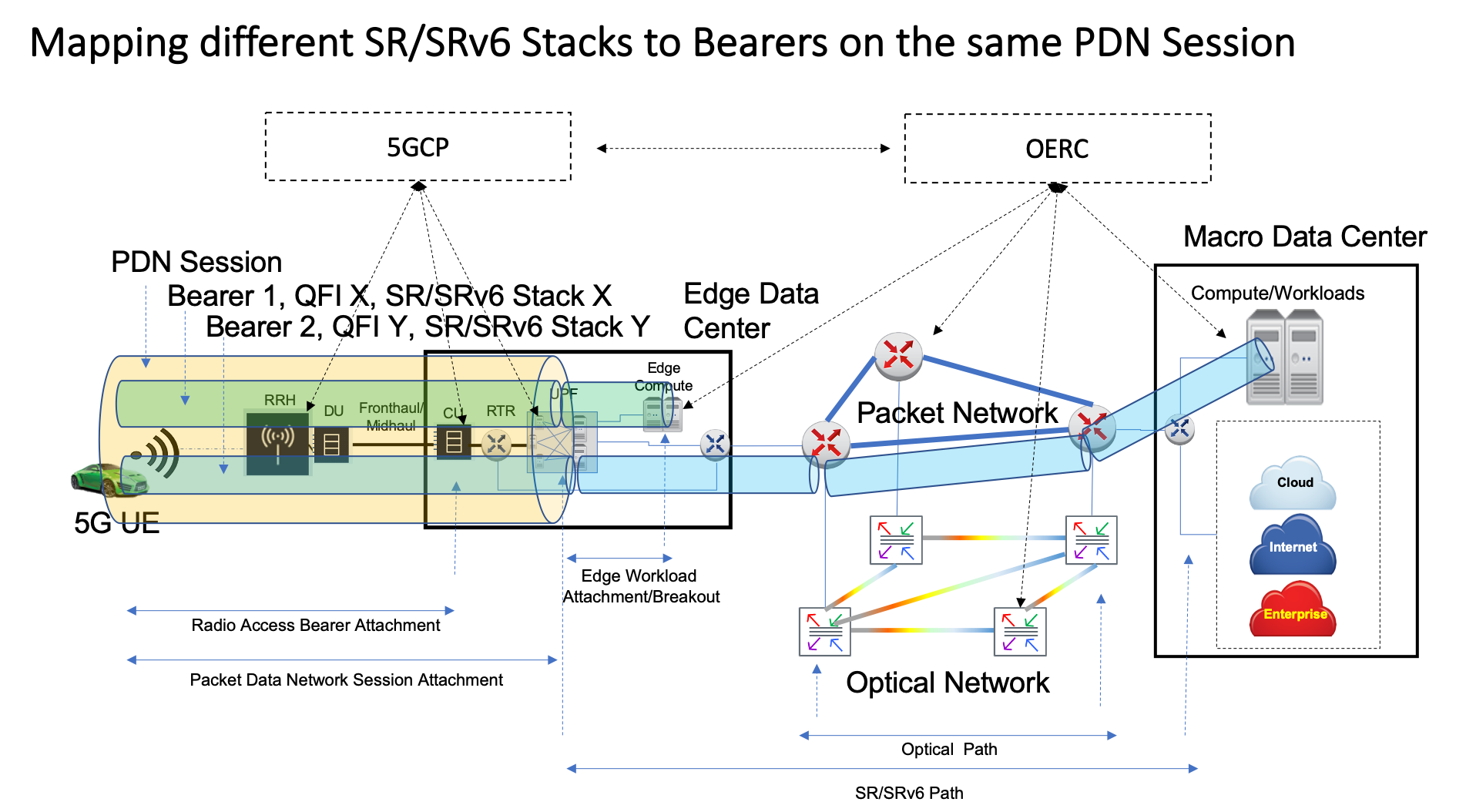}
    \caption{OER with using multiple PDN Bearers mapped to different SR/SRv6 stack on the same PDN Session.}
    \label{fig:oer-qfi}
\end{figure}

Note that the proposed procedure also enables mapping between the 5G QFI provided/provisioned for the UE by the PCF and the traffic class that needs to be transported in the EDC/packet/optical network (the MECD Fabric) on the best path computed by the OER within the MEC Domain.
\begin{itemize}
\item This procedure can be repeated for any required QFI, for example the QFI associated with the traffic Class 1 in our example (or any other defined QFI/Class combination)
\item This procedure represents a possible way of constructing or extending 5G network slices across the EDC/packet/optical network
\end{itemize}
The significance of the ability to map the 5G QFI to SR/SRv6 traffic class in the MECD is that packets from the same UE PDN session mapped by the 5GC to different 5GC PDN bearers corresponding to different QFI values, may be assigned different SR/SRv6 label stacks in the MECD and therefore routed to/from different EC resources. For example,  for the same UE, 5G QFI value X on 5GC Bearer 1 can be mapped to SR/SRv6 Stack X for routing traffic to a macro/central EC resource, while 5G QFI value Y on 5GC Bearer 2 can be mapped to a different SR/SRv6 Stack Y for routing traffic to an EC resource located closer to the UE radio attachment point. This enables traffic routing to multiple EC resources from the same PDN session (on different bearers). The above considerations are shown in Figure \ref{fig:oer-qfi}.

\subsection*{\raggedright{Mobile Edge Routing}}
\label{subsec:mer}
\begin{figure}[htp]
    \centering
    \includegraphics[width=8.5cm]{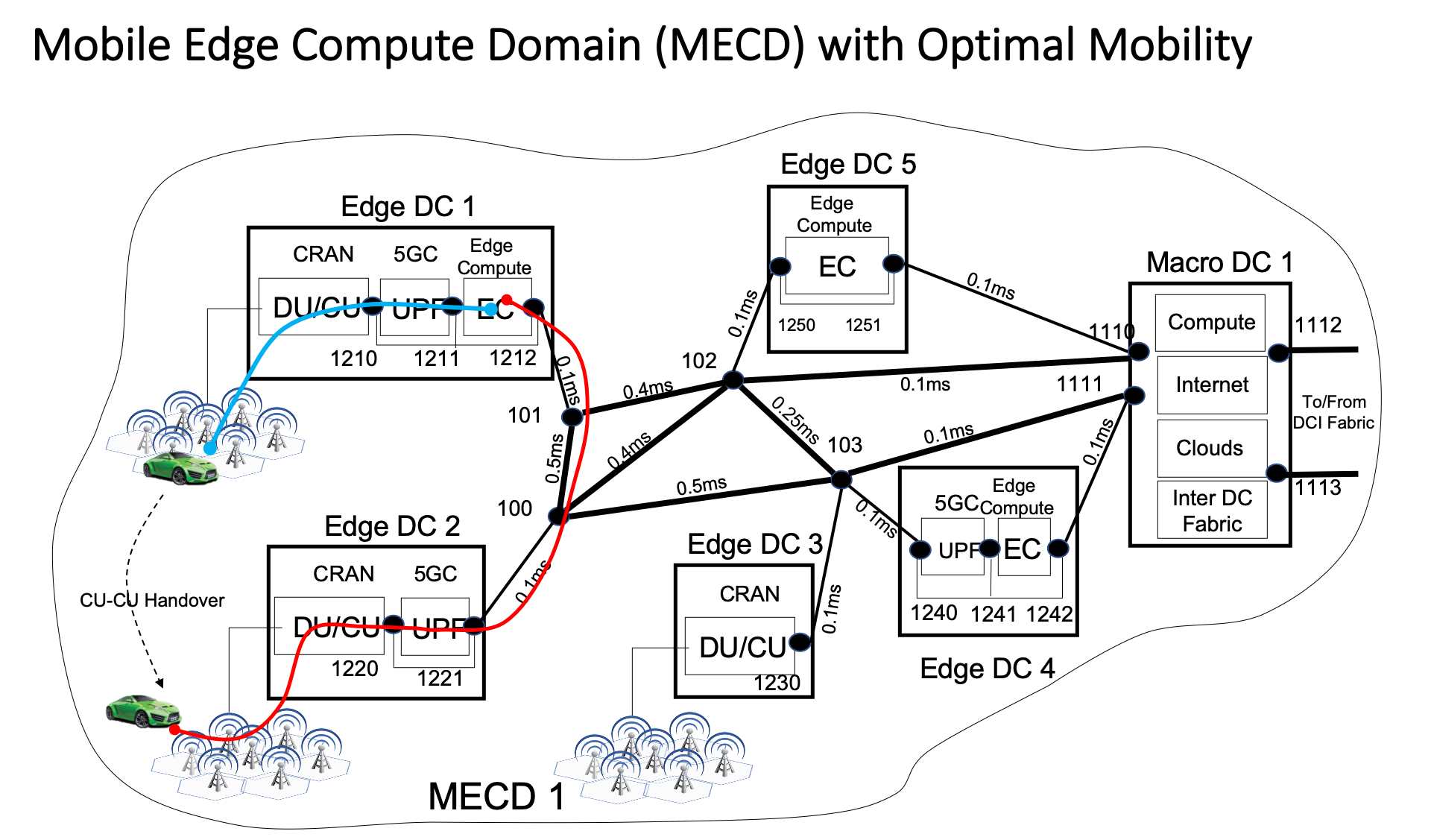}
    \caption{Mobile Edge Routing with inter-CU (gNB) Handover between different EDCs.}
    \label{fig:mer-1}
\end{figure}

We next address the need to maintain Optimal Edge Routing, as defined in Section \ref{subsec:thechallenges}, under mobility conditions. In other words, we wish to ensure that when a UE hands over from a source CU (gNB) in one EDC to a target CU in another EDC within the MECD, the bi-directional traffic path from the target CU to the UPF and EC is maintained within the original traffic class requirements. This is illustrated in Figure \ref{fig:mer-1}. Note that the updated traffic path may involve the target CU and the old or a new UPF (depending on the path computation results). The use of SR or SRv6 on both the UPF and the EC simplifies the path switch process and does not need to involve UPF-to-UPF tunneling (the N9 interface) in order to properly reach the UE IP address, since the traffic forwarding from EC to UE is not based on the UE IP address but rather on the SR/SRv6 label stack.
As shown in the Figure \ref{fig:mer-1}, the inter-CU handover results in the UE attaching to a CU in a new EDC while the original active UPF and EC remain in the old EDC. There is a choice in this scenario to maintain the old UPF or switch the UE session to the new UPF. The critical issue is how to perform this switch while maintaining UE’s IP address and optimal traffic routing. While 5G specifications \cite{3GPP23501} define Session and Service Continuity (SSC) modes, SSC Modes 2 and 3 require a change of the UE IP address if a new UPF PDN Session Anchor (PSA) is used which results in disruption of connectivity. In SSC Mode 1 the UPF PSA remains the same for the duration of the UE session, which allows to maintain UE's original IP address, but may result in a sub-optimal traffic path (a.k.a. triangular routing).

To avoid the IP address and connectivity disruption, in conventional mobility management procedures, the old UPF (PSA) still handles the UE’s original IP address and the new UPF is instructed to tunnel the traffic to the old UPF using GTP-U over N9 interface. In \cite{AECC-Edge} the authors propose to use Uplink Classifiers (ULC) to break out the traffic at the new UPF in order to optimize the traffic path. However, this creates a IP routing implication as the UE’s IP address (a /32 in IPv4 or a specific IPv6 prefix assigned to the UE) must now be “leaked” and announced by the new UPF in order for the Downlink traffic to reach the UE correctly. This also creates a need to maintain, signal and manage the ULCs themselves which are usually based on 5-tupples. This requirement for ULC management across a dynamic set of possible UPFs adds to the complexity at the control, management and provisioning levels.

With SR/SRv6, both the UL traffic from the new UPF to the EC resource and the DL traffic from the EC resource to the UE (via the new UPF) is not routed based on the UE’s IP address but is rather based on the SR/SRv6 labels. This is significant because the Downlink traffic from the EC resource to the UE can follow the SR/SRv6 label stack to the new UPF without the need for UPF-UPF tunneling or UE IP address leaking arrangements.
 
The elimination of tunneling for the new path between the new CU and the new UPF and the old EC, shown in red in Figure \ref{fig:mer-1}, helps route traffic in such a way as to preserve the original traffic class requirements in order to maintain the desired quality of service, for example, by eliminating an extra UPF hop/processing delay.

\begin{figure}[htp]
    \centering
    \includegraphics[width=8.5cm]{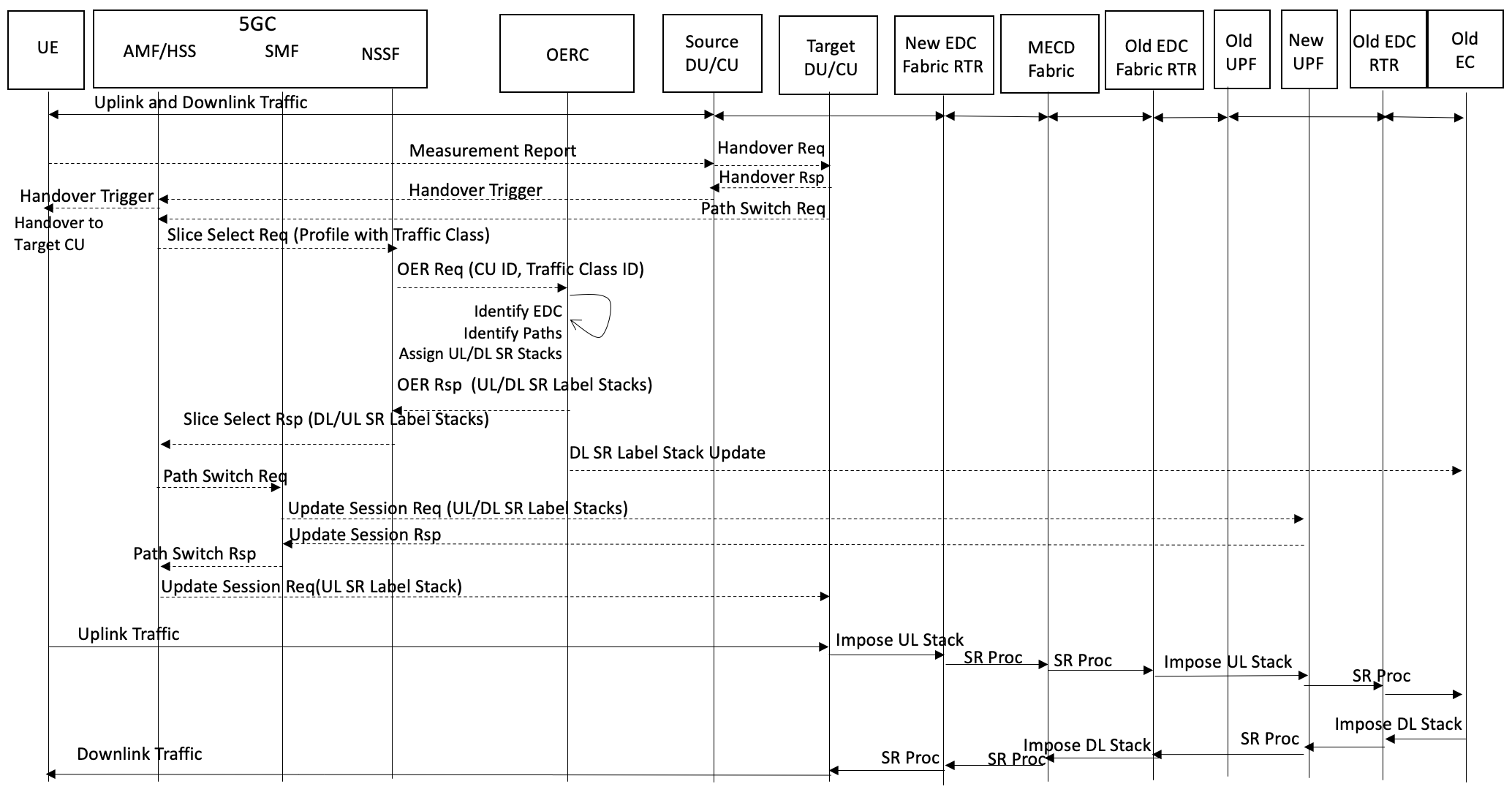}
    \caption{MER Handover and Path Switch signaling procedure.}
    \label{fig:mer-sig}
\end{figure}

\begin{figure}[htp]
    \centering
    \includegraphics[width=8.5cm]{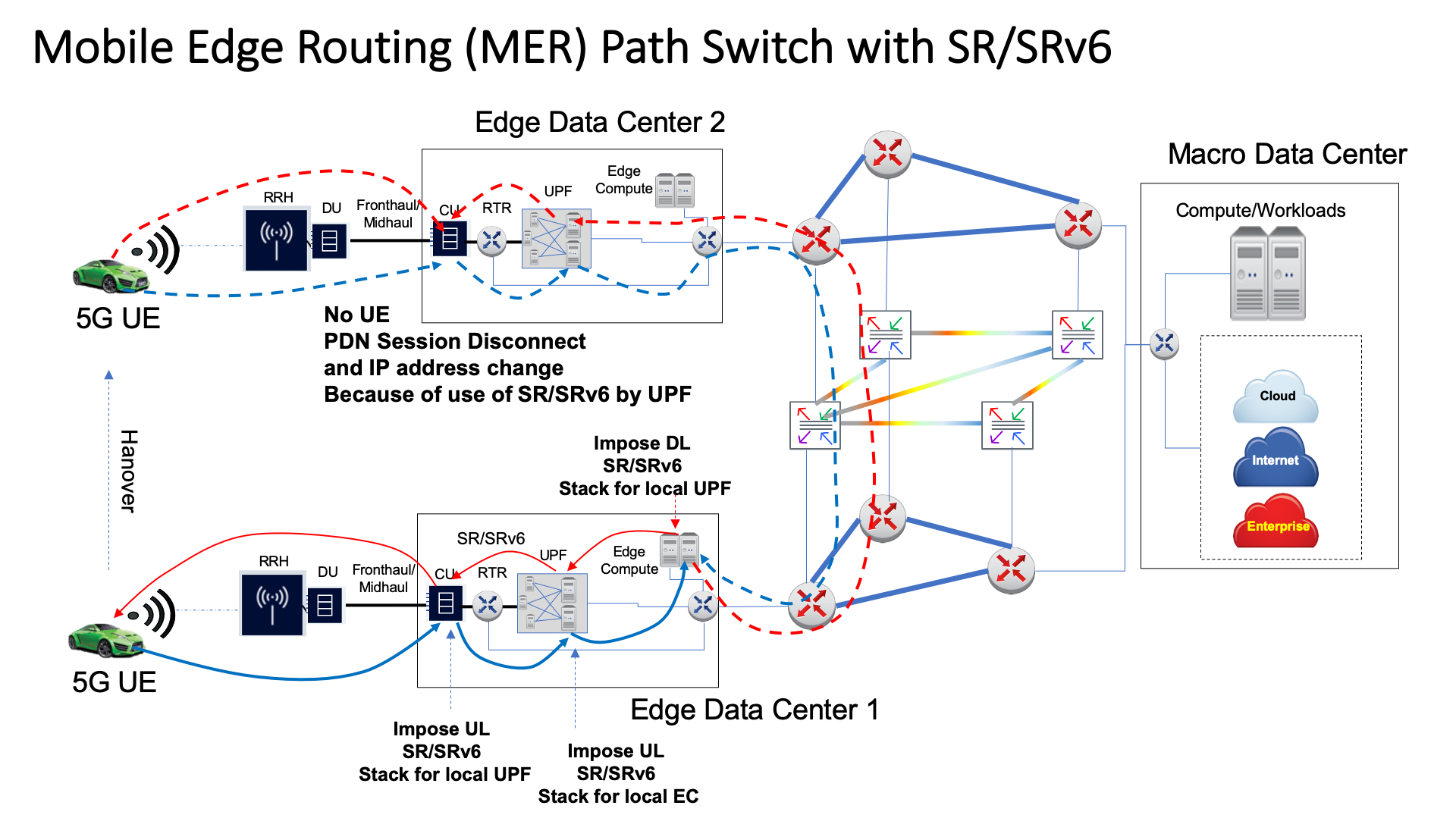}
    \caption{MER Handover and Path Switch using SR/SRv6.}
    \label{fig:mer-path-switch}
\end{figure}

The following procedure is proposed to enable Mobile Edge Routing that involves a “hitless” switch from the old to the new UPF without the need to reconnect the UE and change its IP address:
\begin{enumerate}
\item The UE submits a Measurement Report to the Source CU indicating the Handover (HO) condition.
\item The Source CU signals a HO Request to the Target CU.
\item The Target CU sends a HO Trigger to the Source CU.
\item The Source CU sends a HO Trigger to the AMF. The AMF instructs the UE to attach to the Target CU.
\item The UE attaches to the Target CU. At this time the Downlink (DL) traffic is still directed along the old path via the old UPF toward the Source CU. The Source CU buffers the DL traffic for the UE. The Source CU forwards the DL traffic to the Target CU on the Xn (gNB-gNB) interface. This is not shown in the signaling diagram.
\item The Target CU sends a Path Switch Request to the AMF.
\item The AMF sends a Slice Select Request to the NSSF indicating the new CU ID and the Traffic Class/parameters.
\item The NSSF sends the OER Request to the OERC asking to update the path with new SR labels for the best path satisfying the traffic class requirement between the new Target CU in the new EDC and the old EC.  In this case the OERC identifies the new UPF in the new EDC and the old EC in the old EDC across the MECD Fabric.
\item The OERC computes the new path and responds to NSSF with the new UL and DL label stacks.
\item The NSSF signals to the AMF the new SR paths and labels.
\item The AMF sends the Path Switch Request to the SMF.
\item The SMF updates the new UPF.
\item The AMF updates the new Target CU with the new UL SR label stack.
\item The Target CU imposes the UL label stack on the packets outgoing on the N3 interface using SR or SRv6.
\begin{itemize}
\item The packets reach the new UPF.
\item The new UPF imposes SR/SRv6 label stack on the outgoing packets.
\item The EDC Fabric Router and the MECD Fabric Routers forward the UL packets to the old EDC. 
\item The old EDC routers forward the packets to the EC and the Application Workload.
\end{itemize}
\item In the DL direction:
\begin{itemize}
\item The EC networking stack imposes the DL SR/SRv6 label stack on the packets toward the new UPF via the MECD Fabric, regardless of the UE’s IP address.
\item The MECD Fabric routers forward the DL packets to the new EDC.
\item The new EDC Fabric Router forwards DL packets to the new UPF.
\item The new UPF imposes the SR/SRv6 DL label stack on the packets outgoing on the N3 interface toward the Target CU.
\item The Target CU forwards DL traffic to the DU and the UE.
\end{itemize}
\end{enumerate}
This procedure is shown in Figure \ref{fig:mer-sig}.
The MER path switch to the new UPF without the PDN session disconnect and UE IP address change due to the use of SRE/SRv6 on the UPF is shown in Figure \ref{fig:mer-path-switch}. 

\begin{figure}[htp]
    \centering
    \includegraphics[width=8.5cm]{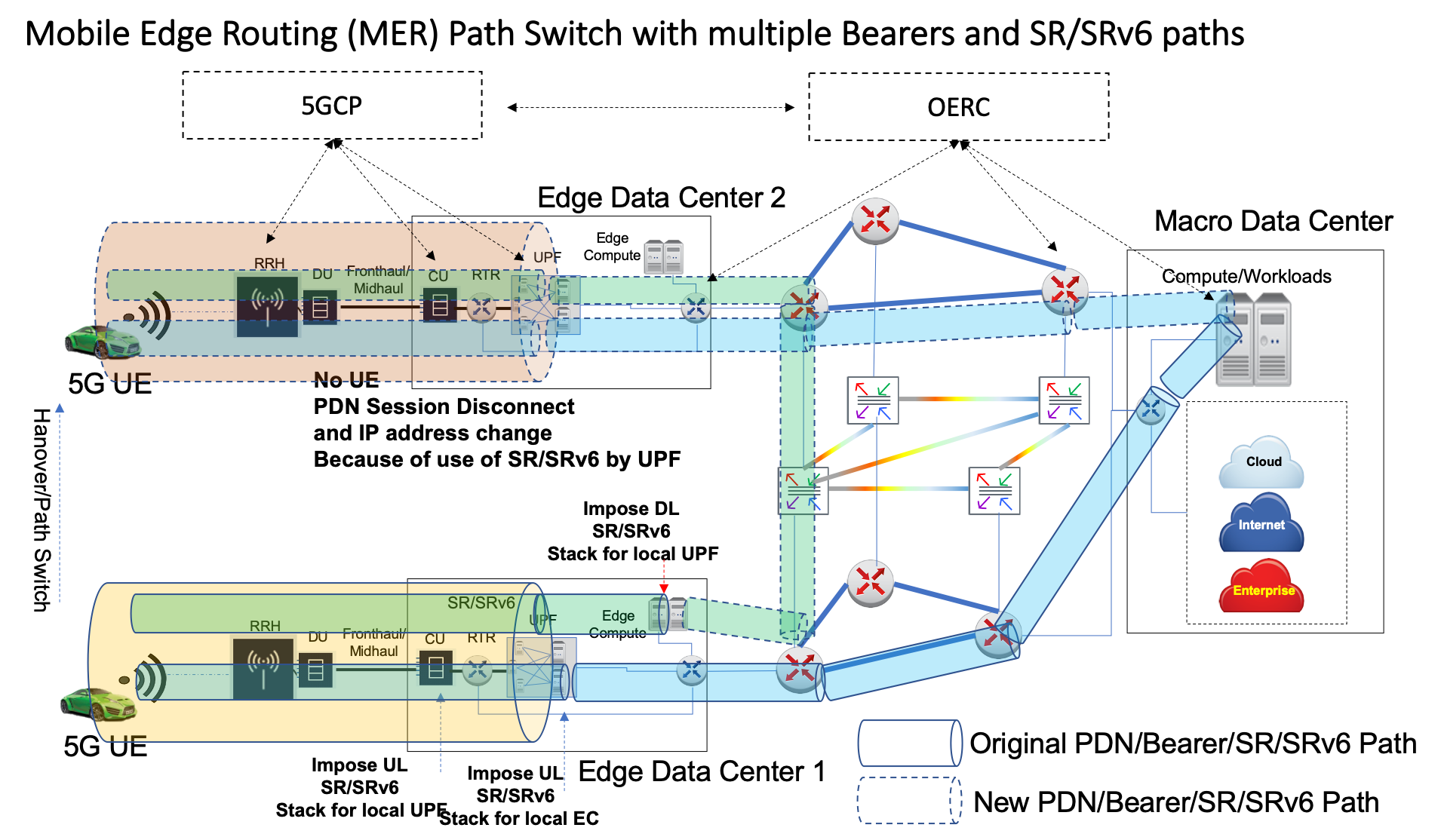}
    \caption{MER Handover and Path Switch with multiple Bearers on one PDN session with SR/SRv6 path mapping to different EC resources.}
    \label{fig:mer-qfi}
\end{figure}

Note that the functional role of the UPFs in this scenario can be compared to the combination SGW-PWG or the SAEGW function in the LTE EPC architecture. The added optimization here is that the UPFs perform this combo function and use SR/SRv6 on both N3 and N6 interfaces without the need for GTP-U. Thus, the switching of UL traffic from the N3 to the N6 interface is based on SR/SRv6 labels only and does not involve the added GTP encapsulation – meaning that this traffic can be directly consumed by the upstream SR/SRv6 capable network and Edge Computing network elements and resources. Similarly, the DL traffic originated by the EC resource using SR/SRv6 can be routed to the correct UPF based on the SR label stack, provided by the OERC, and consumed directly by the UPF on the N6 interface, and then switched to the N3 interface after the SR or SRv6 label switch processing.
Figure \ref{fig:mer-qfi} shows a scenario, where the original PDN session has two Bearers mapped to two different SR/SRv6 label stacks for handling two traffic classes on two different EC resources – one in the local EDC and the second one in the Macro DC. The interaction between the 5GCP and OERC results in the PDN and Bearer Path Switch to the new UPF and the coordinated UL/DL SR label stack update mapped to both bearers to route traffic to the original EC resources along the most optimal new paths in the MECD Fabric. Note that the PDN Path Switch takes place without the UE disconnect and the new IP address assignment as discussed in the OER section.

\subsection*{\raggedright{Application Workload Routing}}
\label{subsec:awr}
Building on the Mobile Edge Routing issue (section \ref{subsec:mer}), it is conceivable to envision that the critical application traffic may be negatively impacted by the additional network distance if the application workloads are still anchored at the original EDC and are effectively lagging behind the moving user. It is therefore desirable to “break-out” the critical application traffic locally at the current EDC. However, this local break-out only makes sense if the application workload is also moved or replicated to the current EDC. The AWR should enable coordinating the Local Break-Out (LBO) decision with the ability to replicate the application workload along the most optimal path to the compute resources local to the break-out point.
\begin{figure}[htp]
    \centering
    \includegraphics[width=8.5cm]{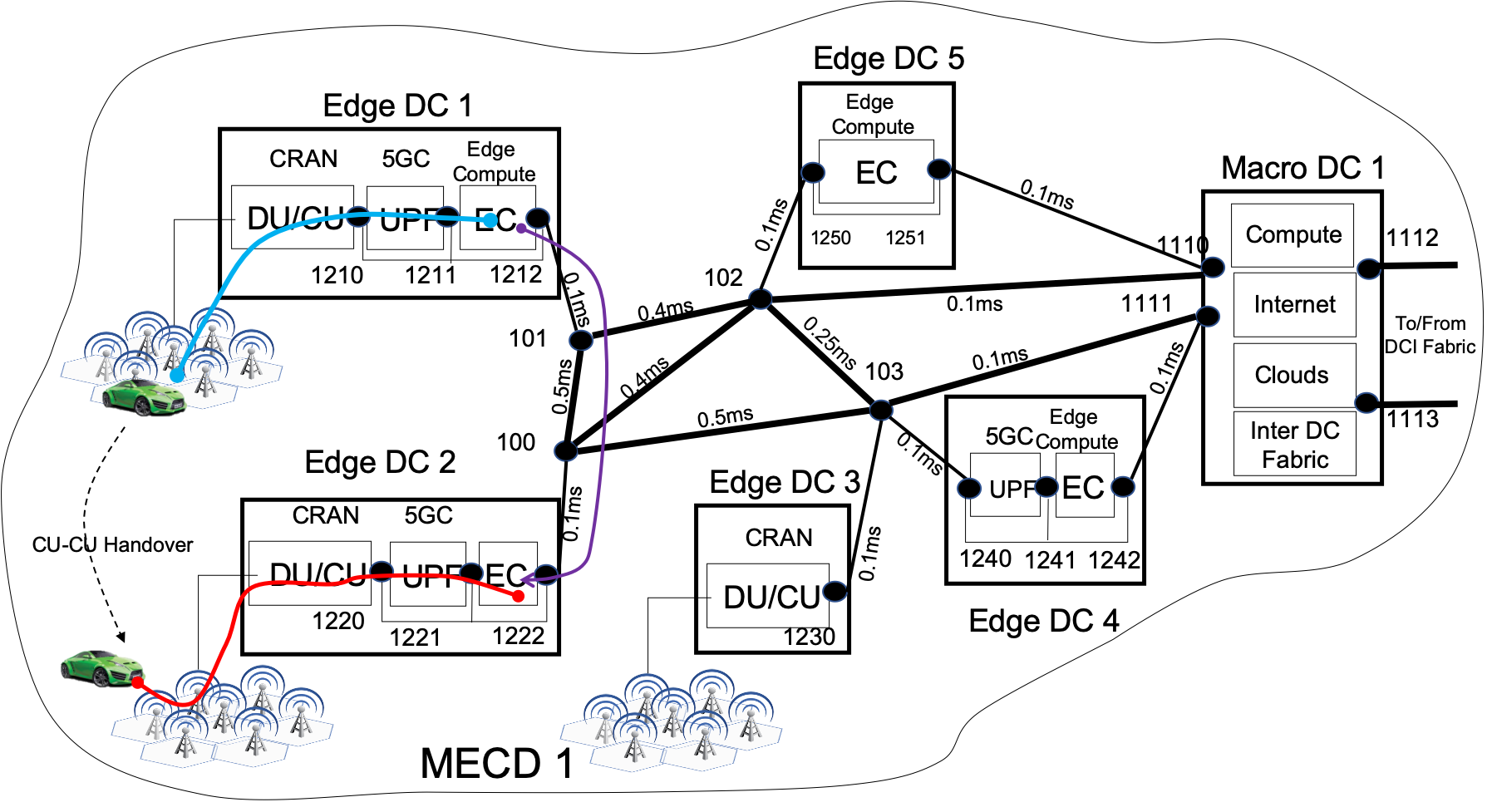}
    \caption{Application Workload Routing under mobility conditions.}
    \label{fig:awr-1}
\end{figure}
Consider the scenario in Figure \ref{fig:awr-1}. A mobile device hands over from the CU in EDC1 to the CU in EDC2. Let us assume that the mobile network has the capability to enable LBO of the critical traffic to the local UPF located in EDC2. Note that the selection of the UPF for the LBO traffic may also be accomplished as part of the MER procedure discussed in section \ref{subsec:mer}. In addition, assume that EDC2 is also equipped with the EC resources that are capable of supporting the application workloads.

While the mobile network LBO capability enables termination of the UL/DL mobile traffic locally in the new EDC, there is an issue with being able to coordinate the application workload or application context (if necessary) replication to the EC resources in the new EDC location such that the traffic class and service requirements can be satisfied. In this case we assume that these requirements are such that the only suitable location for the application workload is the EC resources local to the new EDC. In other words, if the UL/DL traffic has to go back to the original EDC, as shown in Figure \ref{fig:mer-1}, the performance degradation caused by the additional network latency may result in the inability of the application to perform correctly.

One way to deal with this issue is to pre-activate compute resources in all possible locations within the MECD and synchronize the application state as well as the associated data across all these possible locations. This approach, however, may be inefficient and expensive in terms of the complexity of the application software and the reservation of compute resources.

What we are proposing instead, is the ability to inter-work between the mobile network, the OERC and the AWDC to enable coordination of the movement or replication of the application workload on an on-demand basis to the appropriate EC resources, making use of the cloud native capabilities and techniques such as containers and Kubernetes (note that the details of the Kubernetes implementation for this capability are outside of the scope of this paper).
We envision two modes of AWR operation:
\begin{itemize}
\item Reactive Mode. The application workload replication is executed by the AWDC as part of the ongoing signaling sequence between the 5GCP, OERC and AWDC, after the decision to switch to new UPF.
\item Predictive Mode. The application workload replication is executed in anticipation of the mobile device handover to the new EDC based on the run time data learnt from the mobile network and the mobile device itself. For example, the device may supply data to the application workload on its movement pattern, while the mobile network can expose data on the approaching handover event for the device. We can envision a Machine Learning (ML) engine that could consume one or both of these (or any additional) data streams and identify the need to replicate the application workload environment for the device in question in an on-demand manner in the new EDC. This ML engine may be part of the AWDC and may also coordinate the new EC resource location, where the application workload is to be replicated, with the OERC. The OERC, in turn, may influence the 5GCP using the SR/SRv6 label stack methods described earlier.  
\end{itemize}
The Predictive Mode may be based on device data only (i.e. between the application client on the device and the application server in the workload). This may involve device location data, direction of travel, etc.). The data from the mobile network may be used to enhance the application workload distribution decisions. This may present a new monetization opportunity for the mobile network providers.

Assuming that the initial device (UE) connectivity, UL/DL traffic routing and application workload placement is performed according to the procedure described in section \ref{subsec:oer}, and the basic handover procedure (without the application workload replication) is performed as described in section \ref{subsec:mer}, the following figures show the high-level AWR operation in Reactive and Predictive Modes.

\begin{figure}[htp]
    \centering
    \includegraphics[width=8.5cm]{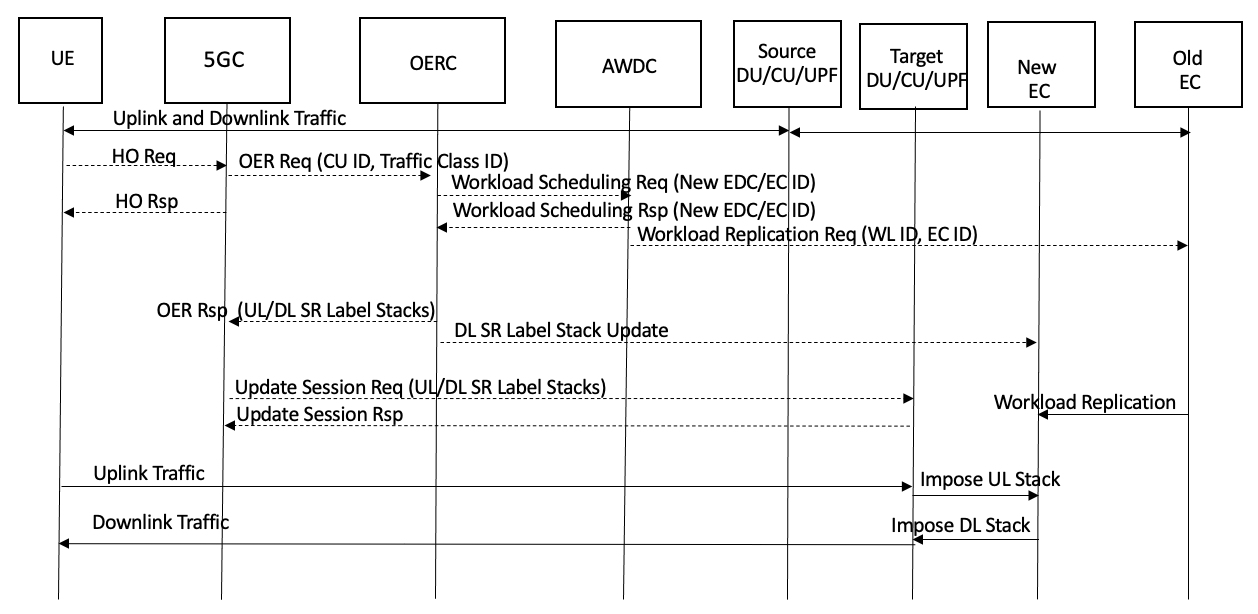}
    \caption{Application Workload Routing in Reactive Mode.}
    \label{fig:awr-rm}
\end{figure}

As shown in Figure \ref{fig:awr-rm}, in the Reactive Mode, the AWDC performs application workload replication during the handover and SR/SRv6 label stack update procedures.

\begin{figure}[htp]
    \centering
    \includegraphics[width=8.5cm]{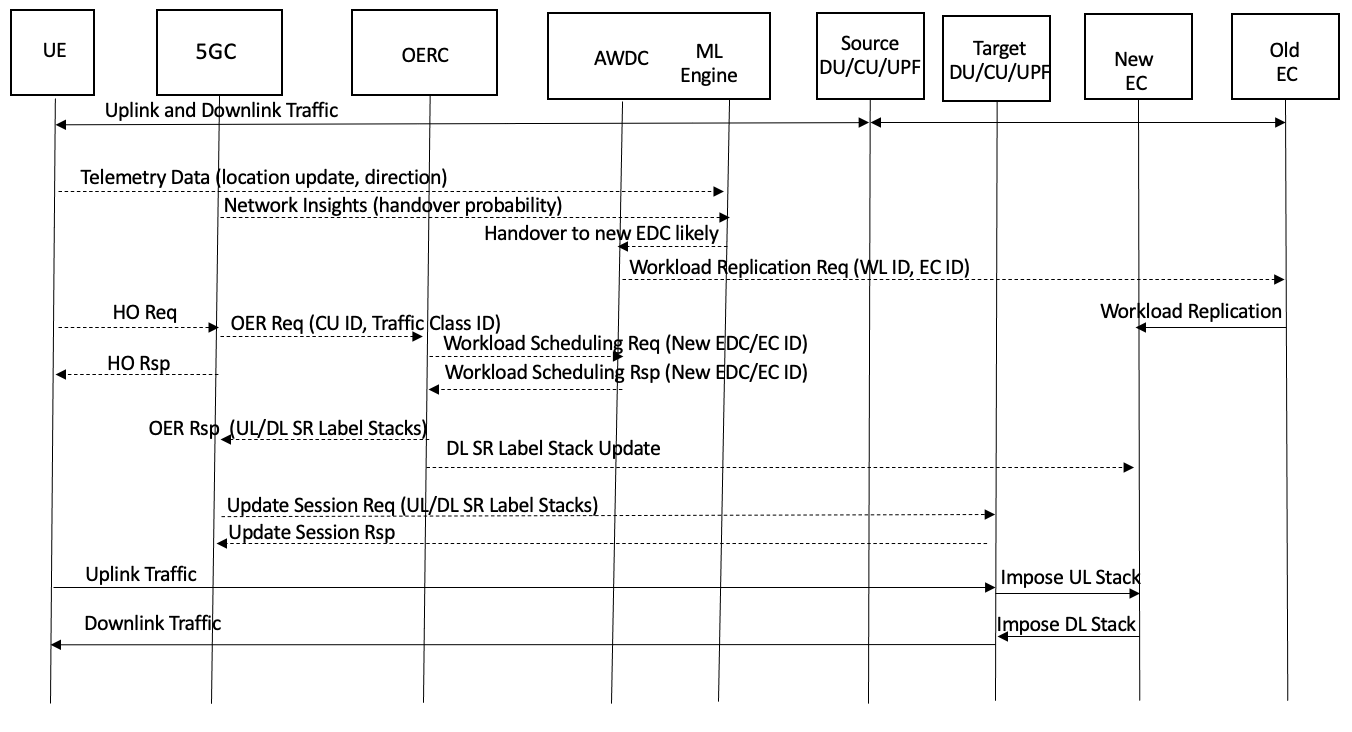}
    \caption{Application Workload Routing in Predictive Mode.}
    \label{fig:awr-pm}
\end{figure}

\begin{figure}[htp]
    \centering
    \includegraphics[width=8.5cm]{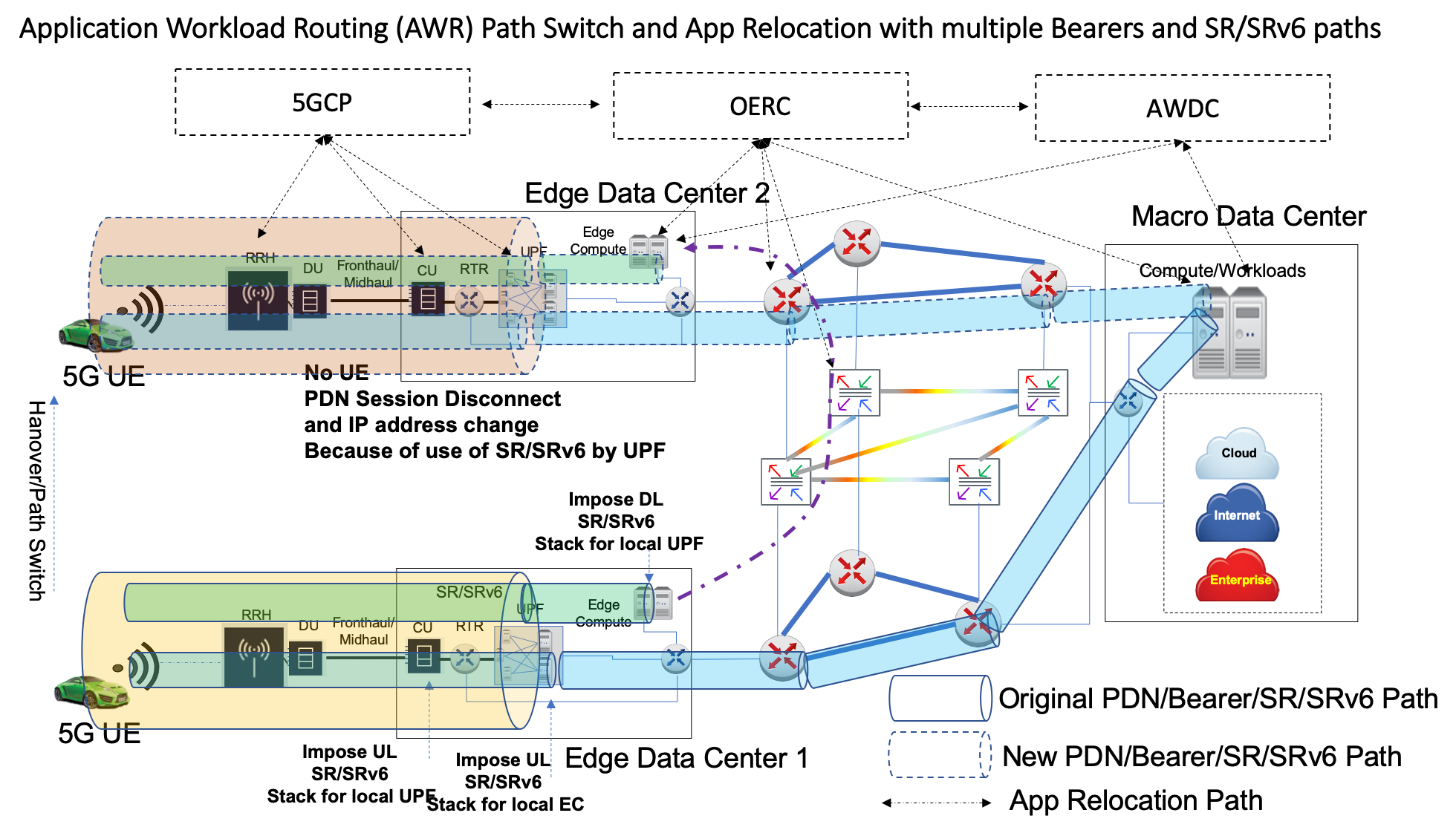}
    \caption{Application Workload Routing with UPF Path Switch and Application Relocation.}
    \label{fig:awr-qfi}
\end{figure}

In Predictive Mode, shown in Figure \ref{fig:awr-pm}, the AWDC makes an early decision to replicate the application workload in the new EDC/EC based on the UE/application telemetry data as well as the insights provided by the mobile network. Please note the use of the word “insights” to reflect a point that the mobile network providers are not likely to share raw data from the network and would rather formulate the insights reflecting the likelihood of events based on the network data.
AWR Path Switch and Application Relocation with multiple bearers and SR/SRv6 is shown in Figure \ref{fig:awr-qfi}.

\subsection*{\raggedright{Application Workload Inter-working}}
\label{subsec:awi}
The Application Workload Inter-working (AWI) addresses the need of enabling 5G mobile devices, operating on the networks from different mobile providers, to make use of common Edge Compute (EC) resources in order to coordinate processing for applications that require sharing of information across multiple network providers. This requirement is compounded by the fact that the application requirements may involve latency, reliability, data rate and other constraints on network connectivity and responsiveness. In addition, the inter-working, application and networking constraints must be satisfied under the mobility conditions as described in the Mobile Edge Routing and Application Workload Routing sections (\ref{subsec:mer}).
\begin{figure}[htp]
    \centering
    \includegraphics[width=8.5cm]{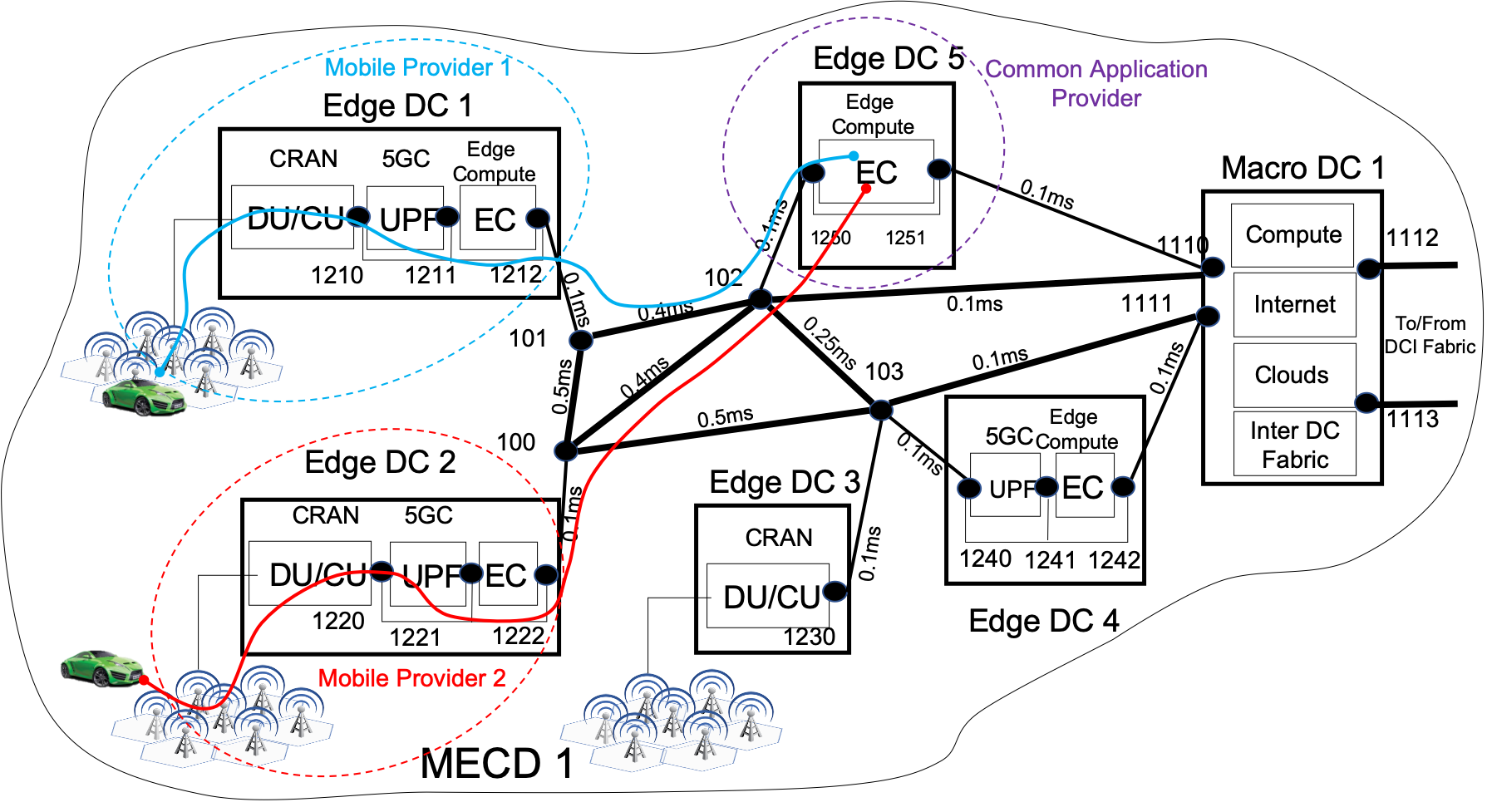}
    \caption{Application Workload Inter-working scenario.}
    \label{fig:awi-1}
\end{figure}
Consider the scenario shown in Figure \ref{fig:awi-1}. A Mobile Network Provider 1 (MNP1) is operating in EDC1. A MNP2 is operating in EDC2. Please note that all elements shown within the EDCs are under the control of the respective MNPs. This includes the RAN, 5GC and EC. Further consider, that the UEs making use of the 5G networks of these two different MNPs must use a common application provided by the Application Provider (AP) operating on Edge Compute in EDC5. We wish to ask the following questions:
\begin{enumerate}
\item How to ensure that the UL and DL traffic from the UEs for the application in question is routed to the common AP workloads within the network performance constraints (e.g. latency, data rate, reliability) imposed by the application?
\item How to ensure the above behavior under mobility conditions?
\end{enumerate}
In order to answer the first question, we can apply methods described in the Optimized Edge Routing section \ref{subsec:oer} with some modifications:
\begin{itemize}
\item The Traffic Class definition may include indicators showing that the Class must use specific or even well-known application resources or IDs. Note that a SR Prefix SID may be used as an application ID. This SID may also be expressed as a SRv6 128-bit address.
\item Edge Compute (EC) element registration may include indicators showing that a specific or a well-known application is present in the EC workloads.
\item The application itself may be associated with specific SR labels (Prefix IDs) or SR Anycast SIDs. This is useful for the application identification and resiliency purposes.
\end{itemize}

\begin{figure}[htp]
    \centering
    \includegraphics[width=8.5cm]{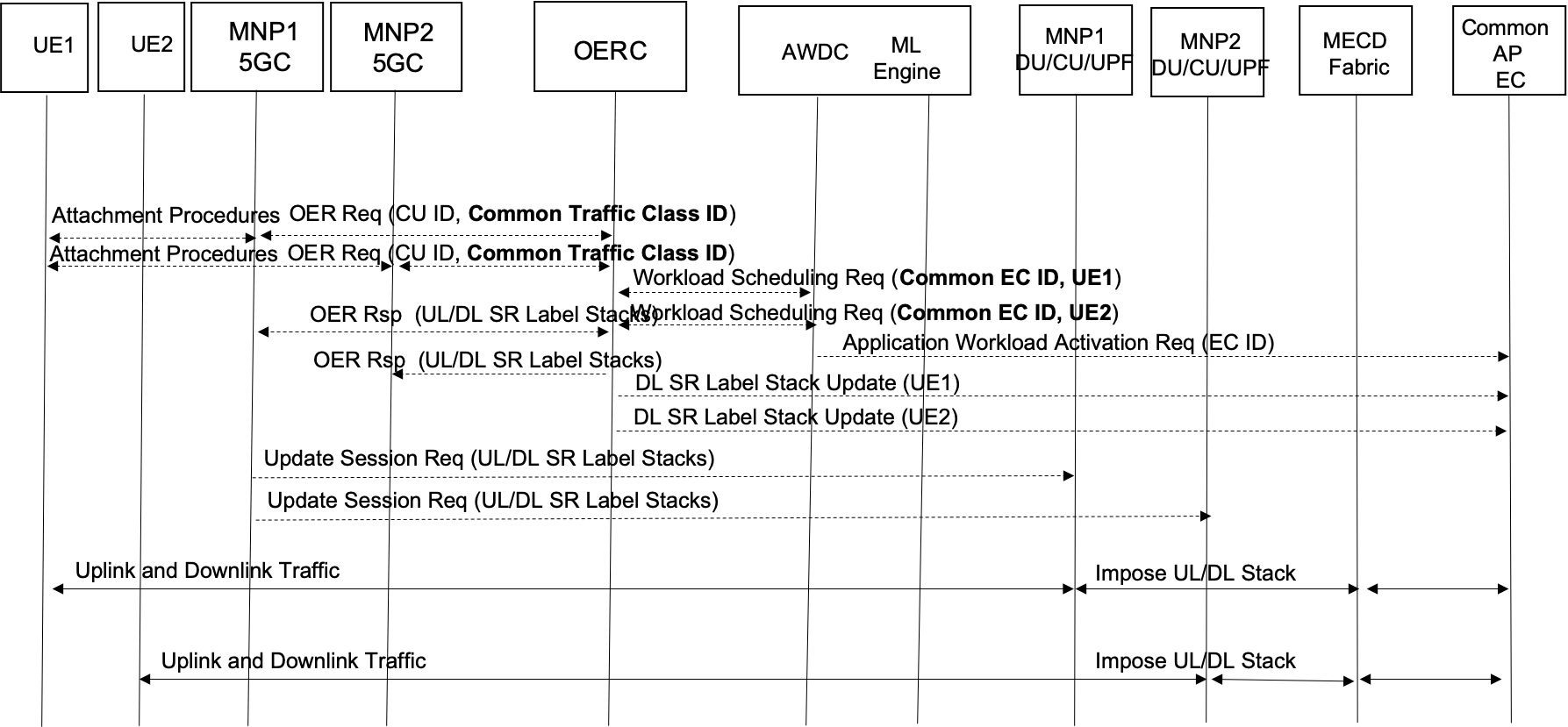}
    \caption{Simplified Application Workload Inter-working signaling procedure.}
    \label{fig:awi-sig}
\end{figure}

A simplified signaling procedure to enable AWI is shown in Figure \ref{fig:awi-sig}.
The second question can be addressed using the same methods as described in the Mobile Edge Routing (MER) section \ref{subsec:mer} and the Application Workload Routing (AWR) section \ref{subsec:awr}. The MER method, with the modifications for the Traffic Class, Edge Compute and Application identification can be used in cases where the replication of the application workload due to mobility is not necessary. The AWR method with the same modifications can be used in cases where there is a need to replicate the common application workload.

\subsection*{\raggedright{Additional Considerations}}
\label{subsec:considerations}
With respect to the Edge Data Offloading and the Edge Compute Server Selection issues described in \cite{AECC-Edge}, the approaches discussed in this paper may be positioned to provide additional or even alternative solutions to both issues.

Specifically, in the 5G network, to avoid the complexities related to managing multiple Uplink Classifiers (ULC) across multiple UPFs, tunneling of traffic between UPFs and making sure that the UE IP address is reachable through all UPFs for the DL traffic, OER, MER and AWR may be applied. The same can be applied to the 4G network with CUPS/LBO, where various traffic classification techniques are used to implement the LBO functionality.

Using OER, the initial Edge Compute resource/server selection and the most optimal, relative to the Edge Compute location, UPF selection/assignment are part of the same Network Slice Selection and OER/MER signaling procedure (please refer to Section \ref{subsec:oer}). Since the Edge Compute resource assignment by AWDC is correlated to the EDC ID where the CU for the UE cellular connection is located, the Edge Compute selection algorithm may be made aware of the UE “topological origination point” (i.e. the location in the EDC network closest to the UE). In addition, since the OERC is aware of the MECD topology and the location of the UPF and EC resources, the OERC may select a UPF that is in turn optimally located relative to the optimal traffic path to/from the EC resource. As indicated in the OER/MER procedures, by returning the SR/SRv6 label stack for the selected UPF to the NSSF, the OERC, and by extension the AWDC, may influence the optimal UPF selection by the 5GCP, thus solving for the Edge Data Offloading issue.

The use of SR or SRv6 by the UPF and the EC (or EC Networking Controller) may eliminate the need for using multiple 5-tupple-based ULCs for traffic break-out in the UL as well as the UE IP address reachability complexities for routing the DL traffic. 

In the UL direction, given that the OERC may assign SR Node SIDs or SRv6 labels/addresses to specific Edge Compute resources or even assign SR Prefix SIDs or SRv6 labels to specific applications associated with these resources, all that the selected UPF needs to do in order to direct the UE UL traffic to the correct Edge Compute or application workload is to impose the SR UL label stack in the outgoing packets (including using SRv6). This is a significant point, because the UE does not even need to be aware of a real IP address of the Edge Compute or the application workload resource in order for the relevant UL traffic to be routed from the UE to Edge Compute. This is because with OER/MER the 5GCP first signals the Traffic Class for the UE session (note that as discussed, the definition of the Traffic Class may also include the application identification and/or a 5G QFI) to the OERC and the OERC returns the UL label stack. As long as the UPF can impose the UL label stack on the packets for the UE PDN connection that is used for this Traffic Class, the upstream MECD Fabric (SR/SRv6 aware) will be able to deliver the packets to intended Edge Compute along the desired path. The use of SR UL stack in this manner can eliminate dependence on IP addresses and the DNS resolution of application URLs to the IPs of application servers or application workloads \footnote{Application workloads are often implemented in Kubernetes pods that share IP addresses of load balancers and require the use of TCP/UDP ports. SR/SRv6 labels can be used to "encode" the current port numbers to simplify reachability to the workloads.}. In other words, the EC/applications may be associated with generic (or anycast) URLs/IPs and the traffic routing to these IPs can be done completely by SR. In addition, the UPF may map the UE UL packets to the SR label stack based on the assigned QFI as opposed to the 5-tupple-based ULC. This capability may be used in multiple ways to:
\begin{itemize}
\item Use a single PDN Session and map multiple Bearers to different SR/SRv6 label stacks to configure end-to-end paths to EC resources located in different places in the MECD topology (e.g. local to an EDC or common/centralized). This mapping may be implemented using 5G QFI or 4G QCI.
\item Extend/map the QoS provided within the 5G/wireless network domain to the QoS provided in the transport and interconnection domains. 
\item Facilitate survivability - the SR Node SID for the Edge Compute resource or a Prefix SID for the application workload may be Anycast SIDs.
\end{itemize}
In the DL direction, since the OERC communicates the SR DL label stack to the Edge Compute resource, and because this stack includes the UPF Node SID, the DL traffic for the UE can be delivered to the correct UPF used for the specific Traffic Class.

In the mobility cases discussed in the MER section \ref{subsec:mer}, if the original UPF and EC assignment does not change, the OERC will find the best path across the MECD Fabric between the new CU for the UE and the original UPF/EC pair. If the UPF is changed as the result of the Path Switch procedures, the use of SR/SRv6 by the EC allows to avoid UPF-UPF tunneling over N9, because the DL traffic follows the DL SR label stack regardless of the UE IP address.

If the LBO functionality (5G or 4G CUPS) is invoked to optimize mobile traffic for the UE in the new location, the combination of the use of the SR label stack by the CU and the ability to assign a more optimal UPF and EC by OERC/AWDC as well as the ability to replicate the application context by AWR/AWI, again solve for both the Edge Data Offload and the Edge Compute selection issues in \cite{AECC-Edge}.

It is important to point out that the User Plane traffic forwarding may be implemented using SRv6 as described in \cite{SRv6-UPF}, in which case the packet forwarding between CU, UPF and EC resources is performed using IPv6 with SR labels encapsulated as 128-bit IPv6 addresses in the SR Header. 

\section{Conclusions}
\label{sec:conclusions}
The presented approaches propose new methods for addressing the issues identified in section \ref{subsec:thechallenges} of this paper. The proposed architecture involving Mobile Edge Computing Domains, the Optimized Edge Routing Controller, the inter-working between the 5G Control Plane, the Application Workload Distribution Control Plane as well as the Edge Computing Networking Control Plane allow to apply Segment Routing or SRv6 traffic engineering and packet forwarding capabilities in an end-to-end manner to optimize mobile traffic routing, eliminate device and server IP addressing dependencies while also optimizing Edge Computing workload placements. 

Optimized Edge Routing within a set of interconnected Edge Data Centers can ensure that traffic requirements in terms of the constraints on latency, throughput, resiliency imposed by the applications are satisfied. This is achieved by utilizing Segment Routing as a fundamental mechanism of end-to-end path identification and traffic forwarding, coupled with methods of element-aware Segment ID registration, identification and inter-working between the 5G Control Plane, Optimized Edge Routing Controller as well as the Application Workload Distribution Controller. The methods proposed in the OER solution can be used to:
\begin{itemize}
\item Influence the User Plane Function selection in the mobile network based on the proximity considerations and path routing within the Edge Compute domain. 
\item Map the Quality of Service parameters within the 5G mobile domain, expressed in terms of the 5G QoS Flow Identifier, to the Traffic Class on the interconnection fabric path between the User Plane Function and Edge Computing resources, expressed in terms of the Segment Routing label stack. 
\item This mapping can be implemented at the Bearer level, allowing using a single PDN Session while associating different SR/SRv6 labels stacks/paths to different Bearers thus enabling per-bearer control for Traffic Class compliant reachability to Edge Compute resources located in different places within the topology.
\end{itemize}
The resulting system allows for intelligent Network Slicing where the routing of traffic in the mobile domain is coordinated with traffic routing in the interconnection domain as well as the application workload placement at the Edge Compute resources within the Mobile Edge Computing domain.

Mobile Edge Routing is applied for mobility scenarios where devices hand over between radio cells that are aggregated in different physical Edge Data Centers. The proposed methods enable optimal traffic routing between the new device point of attachment in the radio network and the original location of the User Plane Function and Edge Computing:
\begin{itemize}
\item Ability to switch UL and DL paths to the new UPF without the need to perform the UPF-UPF tunneling or disconnecting of the UE PDN session and the UE IP address change.
\item Ability to avoid using complex UL traffic breakout methods (such as 5-tupple based 5G Uplink Classifiers) by using SR/SRv6 on the UPF.
\item Ability to avoid complex IP routing scenarios to reach the UE IP address at the appropriate UPF (where the traffic break-out takes place).
\end{itemize}
With Application Workload Routing, the Optimized Edge Routing and the Mobile Edge Routing methods are extended in order to inter-work with the Application Workload Distribution Controller to enable activation and replication of the application workload context in the physical location that is more optimal (e.g. proximal) with respect to the new point of attachment of the mobile device. We specifically propose Predictive Mode that applies Machine Learning utilizing data from the mobile device/application itself as well as the insights from the mobile network in order to proactively replicate the application workload context at the physical Edge Compute resource location where the device is likely to handover.

Application Workload Inter-working applies all previously described methods in order to enable coordination of application processing across multiple mobile network providers and across a set of Edge Computing resources within the Mobile Edge Computing Domain. We propose specific methods of identifying traffic class parameters and Edge Computing resources that indicate the need for common/coordinated processing across multiple mobile network providers.

\section{Acknowledgements}
\label{sec:ack}
The author would like to acknowledge his distinguished colleague Robert J. Huey for numerous and enlightening discussions on practical aspects of designing, deploying and running high-scale networks, Segment Routing, and for his invaluable contribution to this paper.

%\newpage
\bibliography{references} 
%\balance
\end{document}